\documentclass[prb,aps,twocolumn,superscriptaddress,showpacs]{revtex4-1}

\usepackage{graphicx}
\usepackage{epstopdf}
\usepackage[colorlinks=true,linkcolor=black, citecolor=blue, urlcolor=blue,
    unicode=true]{hyperref}

\begin{document}

\title{Spontaneous decay of a soft optical phonon in the relaxor ferroelectric PbMg$_{1/3}$Nb$_{2/3}$O$_{3}$}

\author{C. Stock}
\affiliation{School of Physics and Astronomy, University of Edinburgh, Edinburgh EH9 3JZ, UK}

\author{P. M. Gehring}
\affiliation{NIST Center for Neutron Research, National Institute of Standards and Technology, Gaithersburg, Maryland 20899-6100, USA}

\author{R. A. Ewings}
\affiliation{ISIS Facility, Rutherford Appleton Labs, Chilton, Didcot, OX11 0QX, UK}

\author{G. Xu}
\affiliation{NIST Center for Neutron Research, National Institute of Standards and Technology, Gaithersburg, Maryland 20899-6100, USA}

\author{J. Li}
\affiliation{Department of Materials Science and Engineering, Virginia Tech., Blacksburg, Virginia 24061}

\author{D. Viehland}
\affiliation{Department of Materials Science and Engineering, Virginia Tech., Blacksburg, Virginia 24061}

\author{H. Luo}
\affiliation{Shanghai Institute of Ceramics, Chinese Academy of Sciences, Shanghai, China  201800}

\date{\today}

\begin{abstract}

We report the spontaneous decay of a soft, optical phonon in a solid.  Using neutron spectroscopy, we find that specific phonon lifetimes in the relaxor PbMg$_{1/3}$Nb$_{2/3}$O$_{3}$ are anomalously short within well-defined ranges of energy and momentum.  This behavior is independent of ferroelectric order and occurs when the optical phonon with a specific energy and momentum can kinematically decay into two acoustic phonons with lower phase velocity.  We interpret the well-known relaxor ``waterfall" effect as a form of quasiparticle decay analogous to that previously reported in quantum spin liquids and quantum fluids.

\end{abstract}

\maketitle

\section{Introduction}

Phonons and magnons are elementary excitations that correspond to well-defined deformations of nuclear and magnetic lattices, respectively~\cite{Pines:book}.  These excitations are long-lived for harmonic potentials, but notable cases exist where they spontaneously decay~\cite{Zhit13:85}, resulting in anomalously short lifetimes.  The sudden disappearance of a well-defined, harmonic mode at a particular momentum and energy transfer is rare in lattice dynamics, but it has been reported in several low-dimensional quantum magnets~\cite{Jain17:13,Masuda06:96,Stone06:440,Coldea03:68,Stock15:114} and in strongly correlated electronic systems~\cite{Inosov07:99}.  This process is strictly constrained by kinematics~\cite{Huberman05:72}, and it is more prominent in low-dimensional systems owing to the enhanced phase space for allowed decay routes.  Here we report the spontaneous decay of a soft, optical phonon in a solid.

Relaxors are anharmonic systems that are characterized by a broad, frequency-dependent peak in the temperature dependence of the dielectric permittivity~\cite{Ye98:155}.  The perovskite PbMg$_{1/3}$Nb$_{2/3}$O$_{3}$ (PMN) is arguably the most studied relaxor.  When doped with PbTiO$_{3}$ (PT), PMN and the Zn analogue PbZn$_{1/3}$Nb$_{2/3}$O$_{3}$ (PZN) (PMN-xPT and PZN-xPT), develop exceptional piezoelectric properties that have been exploited for devices~\cite{Park97:82,Service97:275}.  All PMN/PZN relaxors exhibit multiple temperature scales~\cite{Cowley10:60}.  At the Burns temperature T$_{B}$ ($\sim$ 620\,K in PMN), optical refractive index~\cite{Burns09:79} and neutron scattering pair distribution studies~\cite{Jeong05:94} suggest that dynamic, spatially-localized regions of polar order begin to form.  On cooling further to T$_{d}$ ($\sim$ 420\,K in PMN), static short-range polar order appears~\cite{Pirc07:76}, generating intense x-ray~\cite{Vak96:57,You97:79} and neutron~\cite{Vak98:40,Hirota02:65,Hiraka04:70,Gehring09:79} momentum broadened elastic diffuse scattering.  Finally, below the critical temperature T$_{c}$ ($\sim$ 200\,K in PMN)~\cite{Stock07:76}, a transition to long-range ferroelectric order occurs, which in PMN must be induced by cooling in a sufficiently strong external electric field.  These different temperature scales~\cite{Toulouse08:369} can be understood in terms of random field~\cite{Fisch03:67,Pirc01:63} theories, previously applied to model magnets~\cite{Westphal92:68,Stock04:69}.

\begin{figure}
\includegraphics[width=5cm] {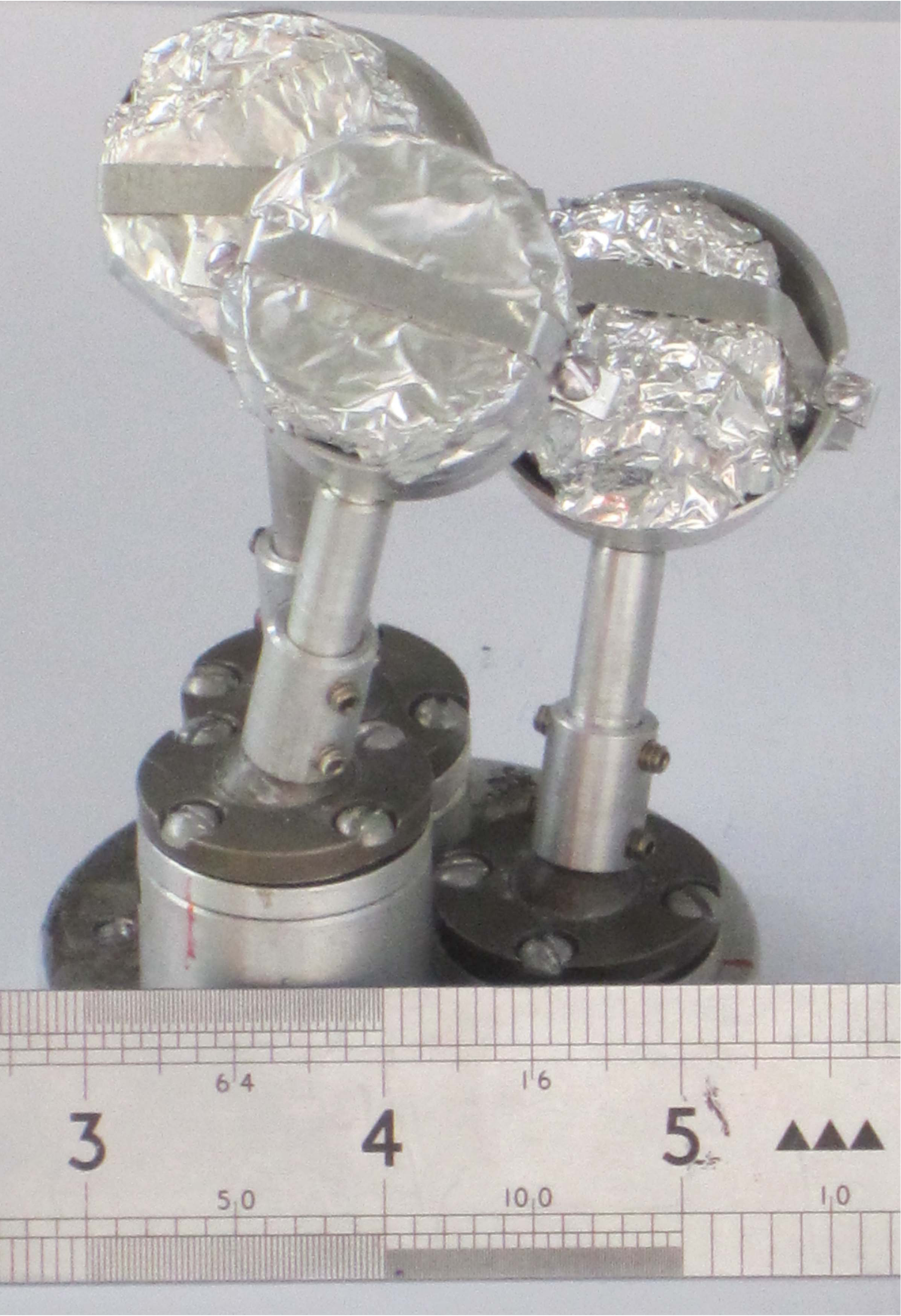}
\caption{\label{crystal} The 192\,g crystal of PbMg$_{1/3}$Nb$_{2/3}$O$_{3}$ used in the neutron scattering experiments, co-aligned and wrapped in aluminium foil.}
\end{figure}

\begin{figure*}
\includegraphics[width=18cm] {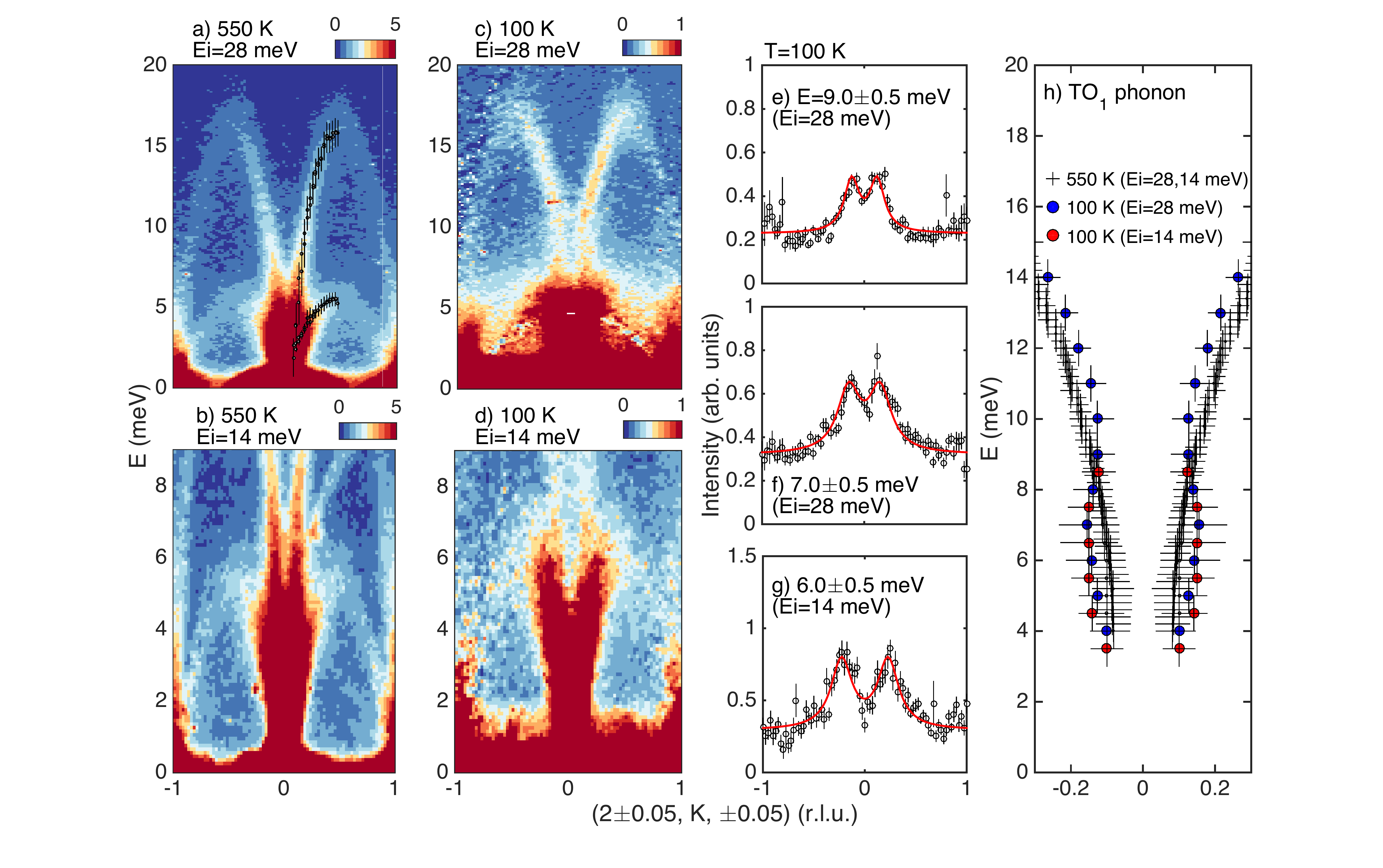}
\caption{\label{merlin_summary} Momentum-energy maps illustrating the waterfall effect at $(a-b)$ 550\,K and $(c-d)$ 100\,K.  $(e-g)$ Constant energy cuts at 100\,K show ridges of scattering (waterfall effect) that persist into the ferroelectric phase.  $(h)$ The $q$ positions of these ridges show no significant difference between 100\,K and 550\,K.}
\end{figure*}

Both PMN and PZN exhibit signatures of strong anharmonicity via the presence of soft phonons at the Brillouin zone center and boundary~\cite{Swainson09:79}.  While early studies failed to identify the soft mode in PMN~\cite{Nab99:11}, subsequent neutron experiments discovered a zone center transverse optic (TO) mode~\cite{Gehring01:87,Wakimoto02:65} near 8\,meV at 1100\,K that softens as T $\rightarrow$ T$_{d}$, as expected from dielectric measurements~\cite{Viehland92:46}, and then hardens to roughly 11\,meV on cooling below T$_{c}$.  The energy of this soft TO mode agrees with Raman scattering at high temperatures~\cite{Taniguchi11:42}, however hyper-Raman studies have identified a lower-energy, zone-center mode that hardens only to $\sim$ 3\,meV on cooling below T$_{c}$~\cite{Zein10:105,Hehlen16:117}.

The neutron scattering lineshape of this higher-energy soft TO mode in lead-based relaxors has an unusual dependence on momentum and energy.  In contrast to systems with temporally well-defined (long-lived) phonons, these compounds display a large temperature region where the TO mode becomes increasingly damped with increasing wavelength~\cite{Gehring00:84}.  This phenomenon was termed the ``waterfall" effect because of the deceptive appearance of an abrupt drop/softening of the TO branch at a non-zero wave vector.  While the waterfall effect was originally tied to the formation of spatially localized polar nanoregions, the phenomena has also been observed in heavily PbTiO$_{3}$-doped materials (in particular PMN-60\%PT) that are not relaxors and show no evidence of polar nanoregions (no diffuse scattering)~\cite{Stock06:73}.  Therefore the waterfall effect is not related to the relaxor nature.

In this paper, we present a momentum resolved study of the waterfall effect in both the low-temperature ``ferroelectric" region and at high temperatures when PMN is paralectric.  We observe anomalous momentum and energy broadened phonon scattering near the Brillouin zone center at all temperatures.  We discuss the phonon lineshape in terms of a decay of the soft optical transverse phonon into lower energy acoustic phonons with lower phase velocities.  We therefore suggest that the waterfall effect is a form of quasiparticle decay analogous to those observed in quantum magnets and fluids.

\section{Experiments}

We reexamine the lattice dynamics of PMN near $q=0$ using the Merlin neutron spectrometer (ISIS, Didcot, UK). Our PMN sample consisted of three crystals, grown using the modified Bridgman technique~\cite{Luo00:39}, that were co-aligned to produce an effective 192 g single crystal (Fig.~\ref{crystal}).  Each sample was aligned such that Bragg reflections of the form (HK0) lay within the horizontal plane.  Experiments were performed on the Merlin direct geometry chopper spectrometer located at the ISIS facility at Rutherford Appleton labs.  The position sensitive detectors on Merlin allowed good momentum resolution to be obtained both within the horizontal scattering plane and also vertically.  The use of position sensitive detectors and the resulting good momentum resolution along all three reciprocal lattice directions distinguishes this measurement from reactor based triple-axis measurements which typically integrate along the vertical direction.  Three different incident energies were used ($E_{i}$=28, 14, and 9\,meV) via the multi-rep rate option with a primary incident neutron energy of 75\,meV and a Fermi chopper frequency of 300\,Hz.  These afforded an energy resolution (full-width at half-maximum) at the elastic line of 1.25\,meV ($E_{i}=28$\,meV), 0.6\,meV ($E_{i}=14$\,meV), and 0.4\,meV ($E_{i}=9$\,meV).

\section{Results}

We first discuss the results from these experiments using the MERLIN spectrometer.  Figure~\ref{merlin_summary} displays intensity maps, measured near $\vec{G} = (200)$ at 550\,K ($>$ T$_{d}$) and 100\,K ($<$ T$_{c}$), that illustrate how the low-energy optical and acoustic phonons disperse for $\vec{q} = \vec{Q} - \vec{G}$.  These phonons correspond to transverse (T$_{1}$) modes propagating along $\pm [010]$ and polarized along [100].  Previous studies using triple-axis spectroscopy were unable to follow the softening of the TO$_{1}$ mode below $\sim 6$\,meV for $|q| < \sim 0.2$\,rlu including $q = 0$ because of significant dampening below the Burns temperature T$_{B}$~\cite{Gehring01:87} that remains until an underdamped lineshape is recovered at T$_{c}$ near 8\,meV~\cite{Wakimoto02:65}.  This is the waterfall effect.  Our neutron TOF data fundamentally modify this picture in two respects:  (1) the soft TO$_{1}$ mode appears to disperse to energies well below 6\,meV in panels $(a-b)$; and (2) there is no evidence of an underdamped TO$_{1}$ mode near 8\,meV at \emph{any} wave vector at 100\,K in panels $(c-d)$.  This finding is significant because it shows that the waterfall effect persists to lower temperatures than previously understood and coexists with ferroelectric order.  The data in panels ($a$) and ($c$) were measured with an incident neutron energy of $E_{i}=28$\,meV, while those shown in panels ($b$) and ($d$) were measured with a much lower value of $E_{i}=14$\,meV.  

The anomalous waterfall feature is characterized further in panels $(e-g)$, which depict constant energy cuts at 100\,K that reveal two peaks in the scattering intensity located symmetrically about $q=0$.  The positions of these peaks are plotted in panel ($h$) from 14\,meV to 3\,meV and demonstrate the existence of a cross-over from a conventional TO${_1}$ phonon dispersion at large $q$ (high energy) to an effectively energy-independent, vertical ridge of scattering below $\sim 10$\,meV.  These vertical ridges of scattering are therefore associated with the waterfall effect.  We also observe no measurable change in the positions of these ridges on cooling from 550\,K to 100\,K.  This response is entirely opposite to that expected for an underdamped, optical phonon branch for which a given wave vector corresponds to a single energy.

\begin{figure}
\includegraphics[width=8.1cm] {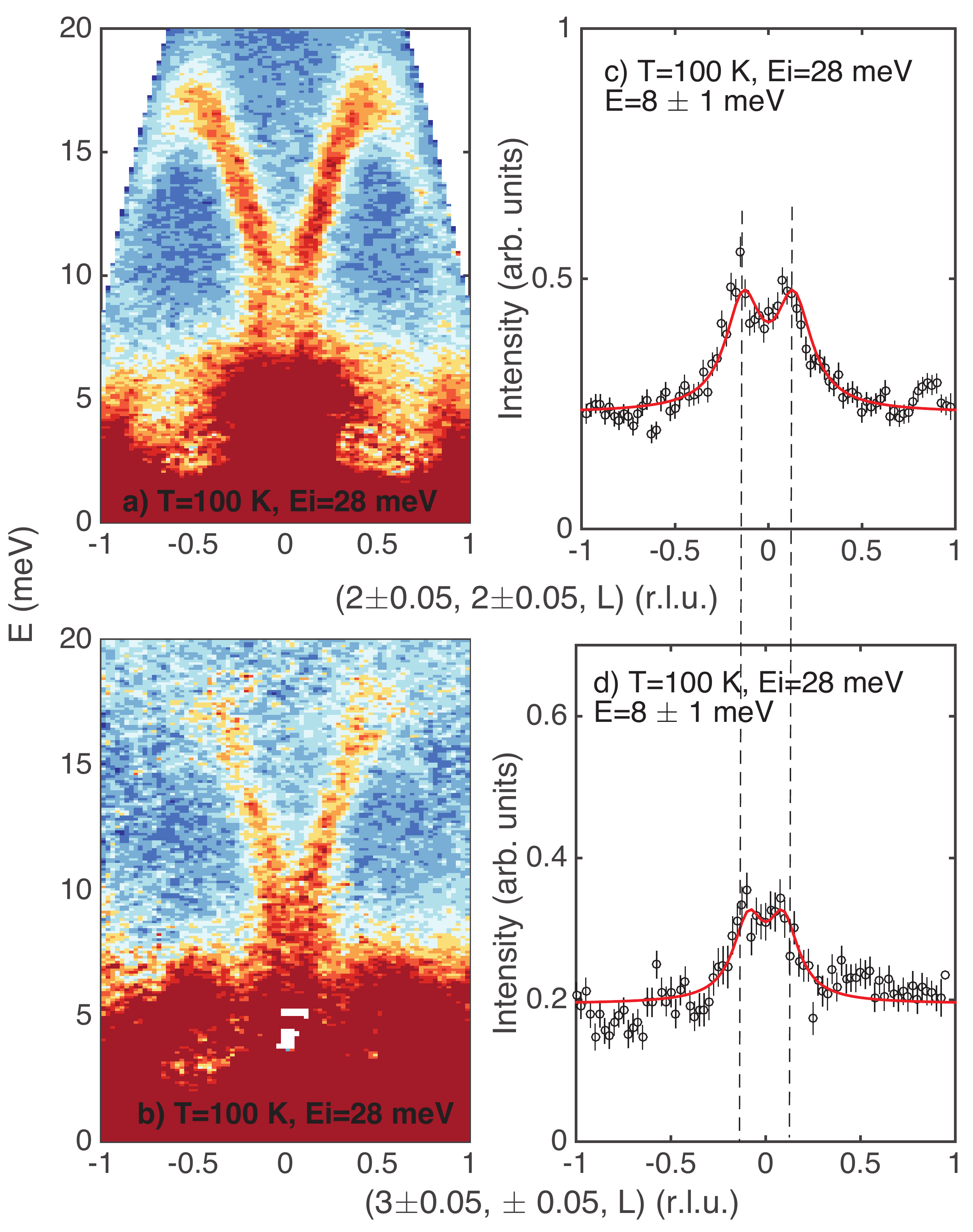}
\caption{\label{zones}  Comparison of the TO$_{1}$ modes measured in the $\vec{Q}=(220)$ and $\vec{Q}=(300)$ zones at 100\,K.  $(c-d)$ Constant energy cuts at 8\,meV.}
\end{figure}

We next discuss the Brillouin zone dependence of the transverse acoustic (TA) mode, as it was previously noted~\cite{Hlinka03:91,Wakimoto02:66} that the TA phonon dispersions measured in the (300) and (200) zones look different.  There are two explanations for this:  (1) the TA and TO phonon modes are coupled~\cite{Hlinka03:91,Wakimoto02:66}, and (2) the TA phonon branch is coupled to an $E=0$ relaxational mode~\cite{Stock05:74,Stock12:86,Stock10:81}.  Either mechanism will cause the TA phonon measured at (3,$-q$,0) to appear softer than that measured at (2,$-q$,0).  We emphasize that this pertains to the \emph{measured} TA phonon energy and not the bare energy, which is Brillouin zone independent.  The reason why the difference in the TA dispersions matters in the context of the waterfall effect is revealed in Fig.~\ref{zones} $(a-b)$, which displays maps that highlight the TO phonon dispersion in the (220) and (300) zones at 100\,K.  In both cases the TO$_{1}$ phonon branch exhibits distinct vertical ridges of scattering located symmetrically about $q=0$ similar to those observed in the (200) zone.  However, the ridges in the (300) zone lie much closer to the zone center.  Linear cuts through these ridges at a constant energy transfer $E=8$\,meV (panels $c$ and $d$) show that the scattering peaks at L$ = \pm 0.135 \pm 0.005$\,rlu near (220) and at L$ = \pm 0.092 \pm 0.007$\,rlu near (300).  This comparison shows that the position of the waterfall, i.\ e.\ the vertical ridges of scattering, correlates with the measured TA phonon dispersion: in zones where the measured TA dispersion is softer the waterfall effect is concentrated closer to $q=0$.

The momentum dependence of the TO$_{1}$ phonon structure factor is plotted in Fig.~\ref{intensity} $(a)$ throughout the (200) Brillouin zone at 100\,K and 550\,K and was obtained by fitting damped harmonic oscillators to constant-$\vec{Q}$ cuts.  The structure factor is found to be constant throughout the zone until $K \sim 0.2$\,rlu, at which point the TO$_{1}$ phonon intensity begins to decrease rapidly with decreasing wave vector.  The energy width $\Gamma$ (inverse lifetime) of the TO$_{1}$ mode (panel $b$) is constant over the same range of wave vector, but it begins to increase at the same point where the structure factor decreases, which indicates concomitant decreases in both the optical mode lifetime and intensity.  These features are indicative of a TO$_{1}$ phonon instability that exists for a range of wave vectors near the zone center.  Panels $(c-d)$ show cuts at a constant $E=8$\,meV through the vertical ridges of scattering (Fig.~\ref{merlin_summary}) and Lorentzian fits show that the locations of these ridges do not change between 100\,K and 550\,K.  Therefore they are not correlated with ferroelectricity or the presence of static polar nanoregions.

\section{Discussion}

\begin{figure}
\includegraphics[width=8.7cm] {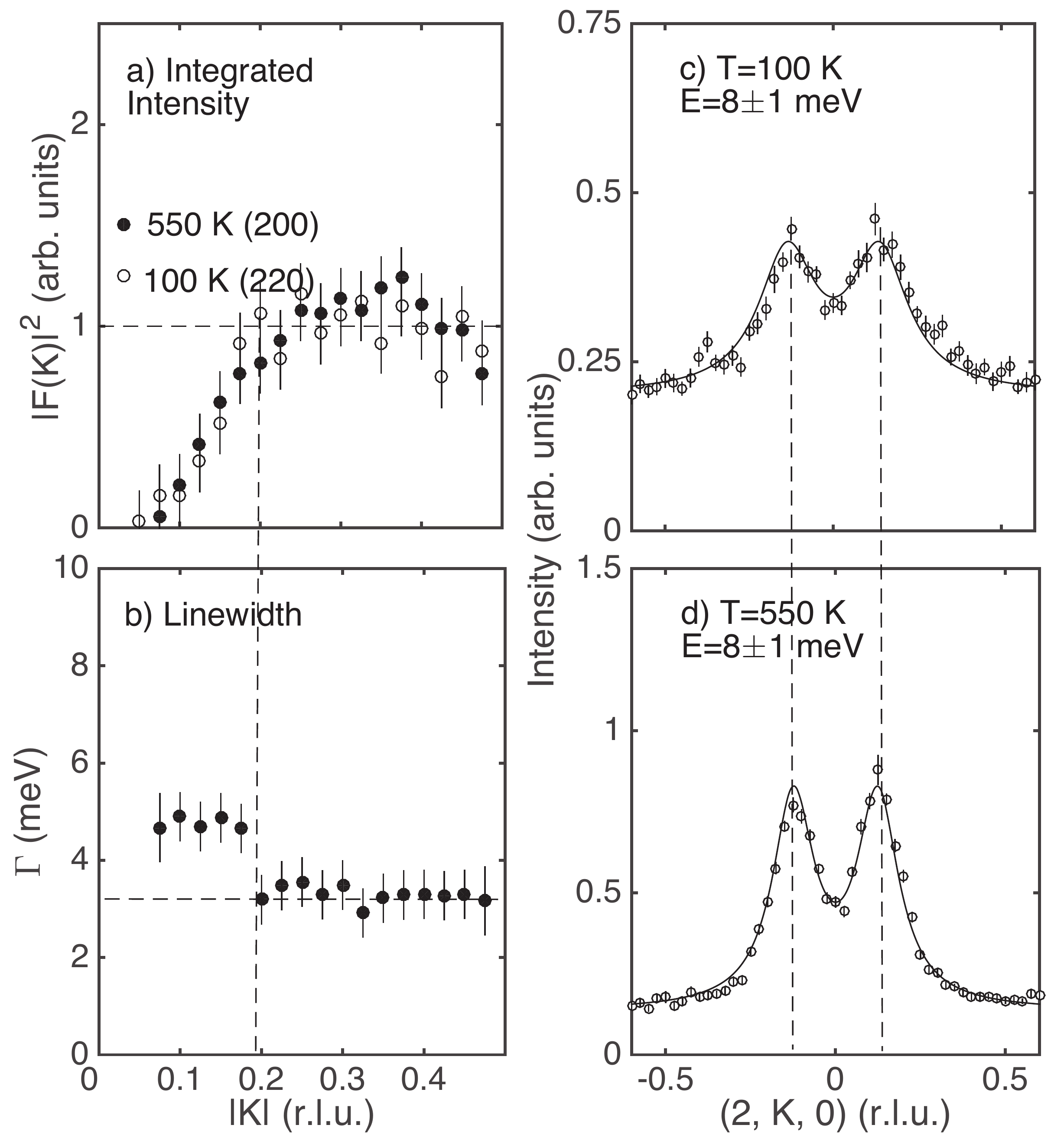}
\caption{\label{intensity} The $q$ dependence of the TO$_{1}$ phonon  $(a)$ structure factor at 550\,K and 100\,K and $(b)$ linewidth.  $(c-d)$ Constant energy $E=8$\,meV scans at 100\,K and 550\,K.}
\end{figure}

The sudden disappearance of a well-defined phonon mode at a particular threshold in momentum and energy is unusual.  However, it has been reported in quantum liquids and magnets.  The energy-momentum broadened nature of the waterfall effect are reminiscent of an energy and momentum scattering continuum, which suggests that it originates from anharmonic processes that involve more than one-phonon scattering.   Higher-order scattering processes, including 3- or 4-phonon processes, are allowed provided that there exists an associated non-zero matrix element and both crystal momentum and energy are conserved.~\cite{Cowley03:68}  We now examine whether or not the decay and loss of spectral weight of the TO$_{1}$ phonon shown in Fig.~\ref{intensity} and the anomalous vertical ridges of scattering shown in Fig.~\ref{merlin_summary} can be explained by a spontaneous decay process analogous to that observed in quantum liquids.

\begin{figure}
\includegraphics[width=9.2cm] {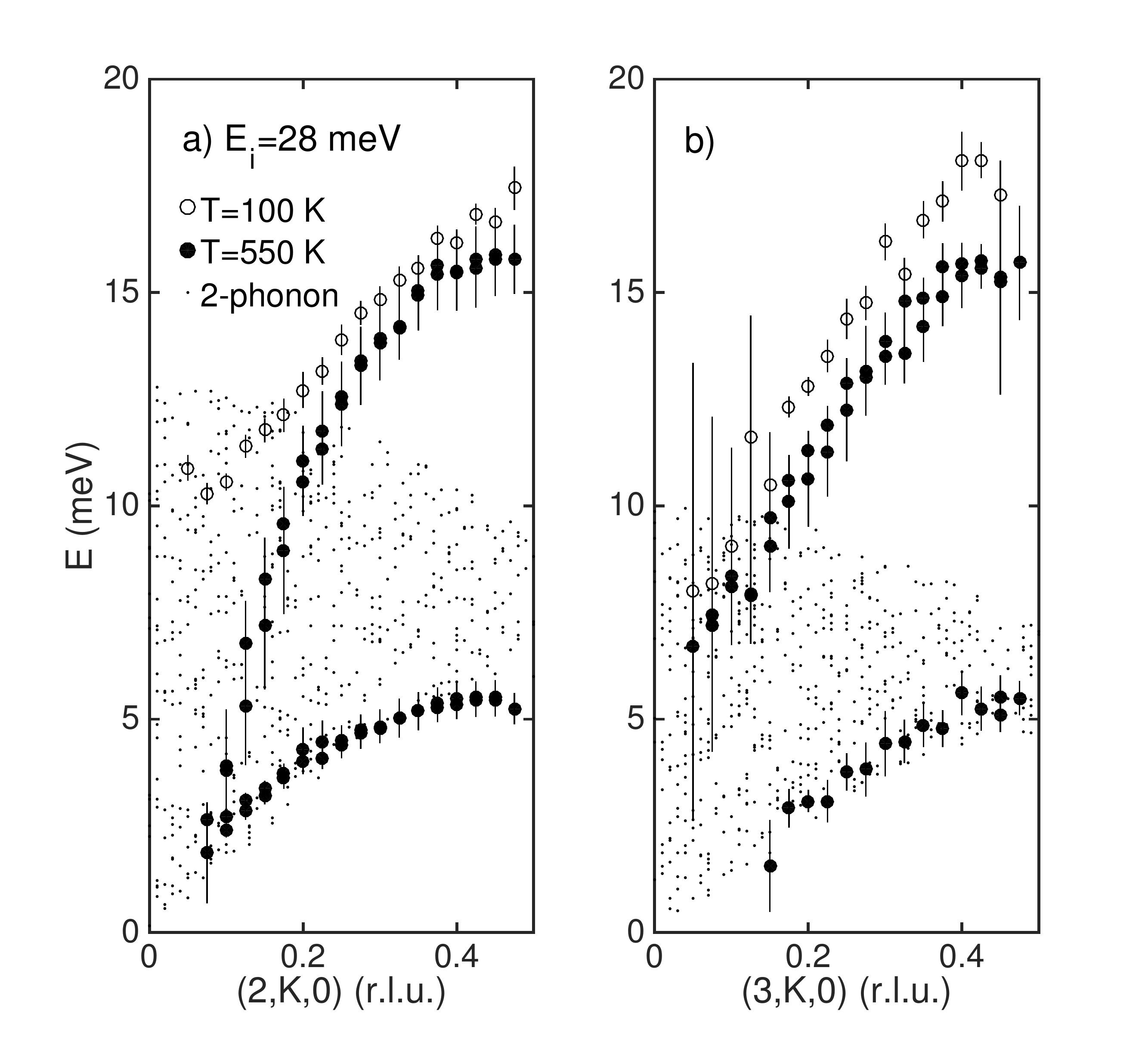}
\caption{\label{dispersion} Dispersions of the TO$_{1}$ and TA$_{1}$ modes from constant-$Q$ cuts with $E_{i}=28$\,meV in the $(a)$ $\vec{Q}=(200)$ and $(b)$ $\vec{Q}=(300)$ zones.  The regions where 2 acoustic-phonon processes are kinematically allowed by from energy/momentum conservation are indicated as discussed in the main text.}
\end{figure}

Panels $(a)$ and $(b)$ of Fig.~\ref{dispersion} show the TA$_{1}$ and TO$_{1}$ phonon dispersions in the (200) and (300) zones, respectively, at 100\,K and 550\,K.  Fig.~\ref{dispersion} also illustrates the kinematically allowed region in each zone where the TO$_{1}$ soft mode (with momentum $\vec{Q}$ and energy $E$) can decay into two TA phonons (with momentum $\vec{Q}_{1,2}$ and energy $\hbar\omega_{\vec{Q}_{1,2}}$) while conserving momentum and energy.  This region is defined by the locus of points where the following expression, forcing both energy and momentum conservation, is nonzero:

\begin{eqnarray}
G(\vec{Q},E)=\sum_{\vec{Q}_{1},\vec{Q}_{2}} \delta(\vec{Q}-\vec{Q}_{1}-\vec{Q}_{2})\delta(E-\hbar\omega_{\vec{Q}_{1}}-\hbar\omega_{\vec{Q}_{2}}) \nonumber
\end{eqnarray}

\noindent In the (200) zone, the $q$ where the soft TO$_{1}$ branch crosses into this kinematically allowed region coincides with that at which the loss of spectral weight and broadening begins (Fig.~\ref{intensity}), namely K$\sim$ 0.2\,rlu.  The corresponding region in the (300) zone is narrower in energy because the measured TA$_{1}$ phonon dispersion is softer.  Kinematically, this implies a narrower decay region centered around $q$=0, confirmed in Fig.~\ref{zones} $(c)$ and $(d)$, which demonstrate that the vertical ridges of scattering (waterfall effect) are concentrated over a narrower range in momentum in the (300) zone than in the (200) Brillouin zone.  Given this agreement with kinematics and the correlation of the waterfall region with the measured acoustic phonon dispersion, we conclude that waterfall effect is due to the spontaneous decay of the soft TO phonon into two TA phonons.

The multi-particle response observed here through the spontaneous decay of a TO mode into two acoustic modes is distinguished by the fact that it occurs at a non-zero threshold in momentum and energy.  This process differs from interference effects between phonons, which have been suggested for PMN and other perovskites and in particular SrTiO$_{3}$ and KTaO$_{3}$~\cite{Yamada69:26,Axe70:1}.  While such phonon interference has been suggested in the context of mode coupling for PMN~\cite{Wakimoto02:66,Hlinka03:91}, an analysis of the temperature dependence of the acoustic phonon energies and intensities has found this explanation to be inconsistent with expectations of acoustic-optic mode coupling based on other perovskites.~\cite{Stock05:74}

The spontaneous decay discussed here can be viewed as being analogous to Fano resonances~\cite{Fano61:124} observed at $\vec{Q}=0$ where an interference can occur between a harmonic mode and a continuum when the energies overlap.~\cite{Struzhkin97:78,Aoki96:76}  In fact, an analogous response has been observed in BaTiO$_{3}$~\cite{ShiraneB70:2} precisely where the optical phonon branch crosses the acoustic branch, resulting in strong interference/coupling effects.  But the spontaneous decay of the optical phonon in PMN occurs in the absence of any such crossing; thus it is not a Fano resonance.  Instead, it is a destructive interference effect that can be observed at non-zero wave vectors only when both momentum and energy are conserved.

Another explanation of the waterfall effect is the coupling between the optic phonon and defects introduced through either disorder or the spatially local ferroelectric order that occurs at low temperatures.~\cite{Ivanov05:47}  The coupling to local ferroelectric order, termed polar nanoregions, is unlikely given that fluctuations associated with this local order freeze in at low temperatures illustrated through both neutron inelastic measurements in an electric field~\cite{Xu08:7,Schneeloch15:92} and also high resolution spin echo results~\cite{Stock10:81,Xu12:86}.  The waterfall effect is present at all temperatures.  Coupling to defects introduced through structural disorder on the $B$ site is likely present at all temperatures.  However, NMR measurements~\cite{Blinc03:91,Laguta03:67} are suggestive that fluctuations associated with this disorder occur on a timescale of $\sim$ kHz, while the optic mode has an energy scale of $\sim$ THz.  These are widely disparate timescales for coupling to occur.  It is also not clear if such a coupling mechanism can explain the momentum dependence between $\vec{Q}$=(200) and (300).

Our results are consistent with Raman spectroscopy~\cite{Taniguchi11:42}, a $lim_{q\rightarrow 0}$ probe, that report highly-damped, high temperature excitations with an energy that agrees with neutrons.  The reported recovery of an underdamped zone-center soft mode at low temperature using hyper-Raman scattering is also consistent with our results because the soft mode energy of 30\,cm$^{-1}$ ($\sim 3.8$\,meV)~\cite{Zein10:105} lies within the resolution of our neutron measurements.  We speculate that the linewidth broadening observed for these modes in Raman measurements may originate from similar multi-phonon processes suggested here.  A continuum of low-energy modes has also been suggested from calculations~\cite{Buss04:16,Takenaka17:546} and diffuse scattering~\cite{Bosak12:68}.

We have motivated our conclusion of spontaneous phonon decay by analogy to quantum spin systems, but such decay processes is not unique to magnetism.  In superfluid $^{4}$He, the interactions between single quasiparticle excitations lead to a breakdown at energies larger than twice the single quasiparticle roton energy.~\cite{Pita59:36,Woods73:36,Shuhl57:1}  This corresponds to a decay of the single roton into multi-particle states.  Also, the decay of longitudinal phonons into two acoustic phonons has been discussed theoretically~\cite{Orbach66:16,Klemens67:28} and used to explain the comparatively subtle phonon broadening ($\sim$1 \%) measured in silicon.~\cite{Klemens66:148}  However, the observation of a complete breakdown of an optical phonon branch has never, to our knowledge, been seen before.  It is even more unusual to observe the decay of a transverse phonon given that they generally have lower phase velocities, however, the softening of the TO mode in PMN near the zone center creates the correct kinematic conditions for a higher energy TO mode to decay into two TA phonons with lower phase velocities.~\cite{Lax81:23}  This effect may be present near other soft-mode driven displacive phase transitions where this condition is met.

Our neutron scattering analysis associates the waterfall effect in PMN with the spontaneous decay of the soft TO mode into multi-particle states based on a comparison of the waterfall effect and the kinematics measured in different Brillouin zones. The origin of this strong decay process is not clear.  The decay rate is linked to anharmonic effects, and such effects may originate from the presence of disorder associated with the Mg$^{2+}$ and Nb$^{5+}$ cations located at the body center of the perovskite unit cell.~\cite{Bose99:271}  The position of the waterfall, i.\ e.\ the wave vector at which the anharmonic scattering is peaked ($q \sim$ 0.1\,rlu), corresponds to fewer than 10 unit cells.  While previous neutron studies have indicated that the temperature scale of the optic mode softening is independent of Ti doping~\cite{Cao08:78}, the momentum position of the vertical ridge of scattering does shift to lower values of $q$ with increasing Ti content~\cite{Gop02:65}, which in turn implies that the length scale associated with this interaction grows.  Anharmonic scattering within such a small length scale frustrates ferroelectric order in the three dimensional perovskite lattice, but relaxing it via Ti doping allows for the eventual development of full ferroelectric order.

\begin{acknowledgments}
This work was supported by the EPSRC, the STFC, the Carnegie Trust for the Universities of Scotland, the Royal Society of London, and the Royal Society of Edinburgh.
\end{acknowledgments}


\begin{thebibliography}{70}%
\makeatletter
\providecommand \@ifxundefined [1]{%
 \@ifx{#1\undefined}
}%
\providecommand \@ifnum [1]{%
 \ifnum #1\expandafter \@firstoftwo
 \else \expandafter \@secondoftwo
 \fi
}%
\providecommand \@ifx [1]{%
 \ifx #1\expandafter \@firstoftwo
 \else \expandafter \@secondoftwo
 \fi
}%
\providecommand \natexlab [1]{#1}%
\providecommand \enquote  [1]{``#1''}%
\providecommand \bibnamefont  [1]{#1}%
\providecommand \bibfnamefont [1]{#1}%
\providecommand \citenamefont [1]{#1}%
\providecommand \href@noop [0]{\@secondoftwo}%
\providecommand \href [0]{\begingroup \@sanitize@url \@href}%
\providecommand \@href[1]{\@@startlink{#1}\@@href}%
\providecommand \@@href[1]{\endgroup#1\@@endlink}%
\providecommand \@sanitize@url [0]{\catcode `\\12\catcode `\$12\catcode
  `\&12\catcode `\#12\catcode `\^12\catcode `\_12\catcode `\%12\relax}%
\providecommand \@@startlink[1]{}%
\providecommand \@@endlink[0]{}%
\providecommand \url  [0]{\begingroup\@sanitize@url \@url }%
\providecommand \@url [1]{\endgroup\@href {#1}{\urlprefix }}%
\providecommand \urlprefix  [0]{URL }%
\providecommand \Eprint [0]{\href }%
\providecommand \doibase [0]{http://dx.doi.org/}%
\providecommand \selectlanguage [0]{\@gobble}%
\providecommand \bibinfo  [0]{\@secondoftwo}%
\providecommand \bibfield  [0]{\@secondoftwo}%
\providecommand \translation [1]{[#1]}%
\providecommand \BibitemOpen [0]{}%
\providecommand \bibitemStop [0]{}%
\providecommand \bibitemNoStop [0]{.\EOS\space}%
\providecommand \EOS [0]{\spacefactor3000\relax}%
\providecommand \BibitemShut  [1]{\csname bibitem#1\endcsname}%
\let\auto@bib@innerbib\@empty
\bibitem [{\citenamefont {Nozieres}\ and\ \citenamefont
  {Pines}(1999)}]{Pines:book}%
  \BibitemOpen
  \bibfield  {author} {\bibinfo {author} {\bibfnamefont {P.}~\bibnamefont
  {Nozieres}}\ and\ \bibinfo {author} {\bibfnamefont {D.}~\bibnamefont
  {Pines}},\ }\href@noop {} {\emph {\bibinfo {title} {The theory of quantum
  liquids}}}\ (\bibinfo  {publisher} {Perseus Books},\ \bibinfo {year}
  {1999})\BibitemShut {NoStop}%
\bibitem [{\citenamefont {Zhitomirsky}\ and\ \citenamefont
  {Chernyshev}(2013)}]{Zhit13:85}%
  \BibitemOpen
  \bibfield  {author} {\bibinfo {author} {\bibfnamefont {M.~E.}\ \bibnamefont
  {Zhitomirsky}}\ and\ \bibinfo {author} {\bibfnamefont {A.~L.}\ \bibnamefont
  {Chernyshev}},\ }\href@noop {} {\bibfield  {journal} {\bibinfo  {journal}
  {Rev. Mod. Phys.}\ }\textbf {\bibinfo {volume} {85}},\ \bibinfo {pages} {219}
  (\bibinfo {year} {2013})}\BibitemShut {NoStop}%
\bibitem [{\citenamefont {Jain}\ \emph {et~al.}(2017)\citenamefont {Jain},
  \citenamefont {Krautloher}, \citenamefont {Porras}, \citenamefont {Ryu},
  \citenamefont {Chen}, \citenamefont {Abernathy}, \citenamefont {Park},
  \citenamefont {Ivanov}, \citenamefont {Chaloupka}, \citenamefont
  {Khaliullin}, \citenamefont {Keimer},\ and\ \citenamefont {Kim}}]{Jain17:13}%
  \BibitemOpen
  \bibfield  {author} {\bibinfo {author} {\bibfnamefont {A.}~\bibnamefont
  {Jain}}, \bibinfo {author} {\bibfnamefont {M.}~\bibnamefont {Krautloher}},
  \bibinfo {author} {\bibfnamefont {J.}~\bibnamefont {Porras}}, \bibinfo
  {author} {\bibfnamefont {G.~H.}\ \bibnamefont {Ryu}}, \bibinfo {author}
  {\bibfnamefont {D.~P.}\ \bibnamefont {Chen}}, \bibinfo {author}
  {\bibfnamefont {D.~L.}\ \bibnamefont {Abernathy}}, \bibinfo {author}
  {\bibfnamefont {J.~T.}\ \bibnamefont {Park}}, \bibinfo {author}
  {\bibfnamefont {A.}~\bibnamefont {Ivanov}}, \bibinfo {author} {\bibfnamefont
  {J.}~\bibnamefont {Chaloupka}}, \bibinfo {author} {\bibfnamefont
  {G.}~\bibnamefont {Khaliullin}}, \bibinfo {author} {\bibfnamefont
  {B.}~\bibnamefont {Keimer}}, \ and\ \bibinfo {author} {\bibfnamefont {B.~J.}\
  \bibnamefont {Kim}},\ }\href@noop {} {\bibfield  {journal} {\bibinfo
  {journal} {Nature Phys.}\ }\textbf {\bibinfo {volume} {13}},\ \bibinfo
  {pages} {633} (\bibinfo {year} {2017})}\BibitemShut {NoStop}%
\bibitem [{\citenamefont {Masuda}\ \emph {et~al.}(2006)\citenamefont {Masuda},
  \citenamefont {Zheludev}, \citenamefont {Manaka}, \citenamefont {Regnault},
  \citenamefont {Chung},\ and\ \citenamefont {Qiu}}]{Masuda06:96}%
  \BibitemOpen
  \bibfield  {author} {\bibinfo {author} {\bibfnamefont {T.}~\bibnamefont
  {Masuda}}, \bibinfo {author} {\bibfnamefont {A.}~\bibnamefont {Zheludev}},
  \bibinfo {author} {\bibfnamefont {H.}~\bibnamefont {Manaka}}, \bibinfo
  {author} {\bibfnamefont {L.~P.}\ \bibnamefont {Regnault}}, \bibinfo {author}
  {\bibfnamefont {J.~H.}\ \bibnamefont {Chung}}, \ and\ \bibinfo {author}
  {\bibfnamefont {Y.}~\bibnamefont {Qiu}},\ }\href@noop {} {\bibfield
  {journal} {\bibinfo  {journal} {Phys. Rev. Lett.}\ }\textbf {\bibinfo
  {volume} {96}},\ \bibinfo {pages} {047210} (\bibinfo {year}
  {2006})}\BibitemShut {NoStop}%
\bibitem [{\citenamefont {Stone}\ \emph {et~al.}(2006)\citenamefont {Stone},
  \citenamefont {Zaliznyak}, \citenamefont {Hong}, \citenamefont {Broholm},\
  and\ \citenamefont {Reich}}]{Stone06:440}%
  \BibitemOpen
  \bibfield  {author} {\bibinfo {author} {\bibfnamefont {M.~B.}\ \bibnamefont
  {Stone}}, \bibinfo {author} {\bibfnamefont {I.~A.}\ \bibnamefont
  {Zaliznyak}}, \bibinfo {author} {\bibfnamefont {T.}~\bibnamefont {Hong}},
  \bibinfo {author} {\bibfnamefont {C.~L.}\ \bibnamefont {Broholm}}, \ and\
  \bibinfo {author} {\bibfnamefont {D.~H.}\ \bibnamefont {Reich}},\ }\href@noop
  {} {\bibfield  {journal} {\bibinfo  {journal} {Nature}\ }\textbf {\bibinfo
  {volume} {440}},\ \bibinfo {pages} {187} (\bibinfo {year}
  {2006})}\BibitemShut {NoStop}%
\bibitem [{\citenamefont {Coldea}\ \emph {et~al.}(2003)\citenamefont {Coldea},
  \citenamefont {Tennant},\ and\ \citenamefont {Tylczynski}}]{Coldea03:68}%
  \BibitemOpen
  \bibfield  {author} {\bibinfo {author} {\bibfnamefont {R.}~\bibnamefont
  {Coldea}}, \bibinfo {author} {\bibfnamefont {D.~A.}\ \bibnamefont {Tennant}},
  \ and\ \bibinfo {author} {\bibfnamefont {Z.}~\bibnamefont {Tylczynski}},\
  }\href@noop {} {\bibfield  {journal} {\bibinfo  {journal} {Phys. Rev. B}\
  }\textbf {\bibinfo {volume} {68}},\ \bibinfo {pages} {134424} (\bibinfo
  {year} {2003})}\BibitemShut {NoStop}%
\bibitem [{\citenamefont {Stock}\ \emph {et~al.}(2015)\citenamefont {Stock},
  \citenamefont {Rodriguez-Rivera}, \citenamefont {Schmalzl}, \citenamefont
  {Rodriguez}, \citenamefont {Stunault},\ and\ \citenamefont
  {Petrovic}}]{Stock15:114}%
  \BibitemOpen
  \bibfield  {author} {\bibinfo {author} {\bibfnamefont {C.}~\bibnamefont
  {Stock}}, \bibinfo {author} {\bibfnamefont {J.~A.}\ \bibnamefont
  {Rodriguez-Rivera}}, \bibinfo {author} {\bibfnamefont {K.}~\bibnamefont
  {Schmalzl}}, \bibinfo {author} {\bibfnamefont {E.~E.}\ \bibnamefont
  {Rodriguez}}, \bibinfo {author} {\bibfnamefont {A.}~\bibnamefont {Stunault}},
  \ and\ \bibinfo {author} {\bibfnamefont {C.}~\bibnamefont {Petrovic}},\
  }\href@noop {} {\bibfield  {journal} {\bibinfo  {journal} {Phys. Rev. Lett.}\
  }\textbf {\bibinfo {volume} {114}},\ \bibinfo {pages} {247005} (\bibinfo
  {year} {2015})}\BibitemShut {NoStop}%
\bibitem [{\citenamefont {Inosov}\ \emph {et~al.}(2007)\citenamefont {Inosov},
  \citenamefont {Fink}, \citenamefont {Kordyuk}, \citenamefont {Borisenko},
  \citenamefont {Zabolotnyy}, \citenamefont {Schuster}, \citenamefont
  {Knupfer}, \citenamefont {B\"uchner}, \citenamefont {Follath}, \citenamefont
  {D\"urr}, \citenamefont {Eberhardt}, \citenamefont {Hinkov}, \citenamefont
  {Keimer},\ and\ \citenamefont {Berger}}]{Inosov07:99}%
  \BibitemOpen
  \bibfield  {author} {\bibinfo {author} {\bibfnamefont {D.~S.}\ \bibnamefont
  {Inosov}}, \bibinfo {author} {\bibfnamefont {J.}~\bibnamefont {Fink}},
  \bibinfo {author} {\bibfnamefont {A.~A.}\ \bibnamefont {Kordyuk}}, \bibinfo
  {author} {\bibfnamefont {S.~V.}\ \bibnamefont {Borisenko}}, \bibinfo {author}
  {\bibfnamefont {V.~B.}\ \bibnamefont {Zabolotnyy}}, \bibinfo {author}
  {\bibfnamefont {R.}~\bibnamefont {Schuster}}, \bibinfo {author}
  {\bibfnamefont {M.}~\bibnamefont {Knupfer}}, \bibinfo {author} {\bibfnamefont
  {B.}~\bibnamefont {B\"uchner}}, \bibinfo {author} {\bibfnamefont
  {R.}~\bibnamefont {Follath}}, \bibinfo {author} {\bibfnamefont {H.~A.}\
  \bibnamefont {D\"urr}}, \bibinfo {author} {\bibfnamefont {W.}~\bibnamefont
  {Eberhardt}}, \bibinfo {author} {\bibfnamefont {V.}~\bibnamefont {Hinkov}},
  \bibinfo {author} {\bibfnamefont {B.}~\bibnamefont {Keimer}}, \ and\ \bibinfo
  {author} {\bibfnamefont {H.}~\bibnamefont {Berger}},\ }\href@noop {}
  {\bibfield  {journal} {\bibinfo  {journal} {Phys. Rev. Lett.}\ }\textbf
  {\bibinfo {volume} {99}},\ \bibinfo {pages} {237002} (\bibinfo {year}
  {2007})}\BibitemShut {NoStop}%
\bibitem [{\citenamefont {Huberman}\ \emph {et~al.}(2005)\citenamefont
  {Huberman}, \citenamefont {Coldea}, \citenamefont {Cowley}, \citenamefont
  {Tennant}, \citenamefont {Leheny}, \citenamefont {Christianson},\ and\
  \citenamefont {Frost}}]{Huberman05:72}%
  \BibitemOpen
  \bibfield  {author} {\bibinfo {author} {\bibfnamefont {T.}~\bibnamefont
  {Huberman}}, \bibinfo {author} {\bibfnamefont {R.}~\bibnamefont {Coldea}},
  \bibinfo {author} {\bibfnamefont {R.~A.}\ \bibnamefont {Cowley}}, \bibinfo
  {author} {\bibfnamefont {D.~A.}\ \bibnamefont {Tennant}}, \bibinfo {author}
  {\bibfnamefont {R.~L.}\ \bibnamefont {Leheny}}, \bibinfo {author}
  {\bibfnamefont {R.~J.}\ \bibnamefont {Christianson}}, \ and\ \bibinfo
  {author} {\bibfnamefont {C.~D.}\ \bibnamefont {Frost}},\ }\href@noop {}
  {\bibfield  {journal} {\bibinfo  {journal} {Phys. Rev. B}\ }\textbf {\bibinfo
  {volume} {72}},\ \bibinfo {pages} {014413} (\bibinfo {year}
  {2005})}\BibitemShut {NoStop}%
\bibitem [{\citenamefont {Ye}(1998)}]{Ye98:155}%
  \BibitemOpen
  \bibfield  {author} {\bibinfo {author} {\bibfnamefont {Z.~G.}\ \bibnamefont
  {Ye}},\ }\href@noop {} {\bibfield  {journal} {\bibinfo  {journal} {Key Eng.
  Mater.}\ }\textbf {\bibinfo {volume} {155-156}},\ \bibinfo {pages} {81}
  (\bibinfo {year} {1998})}\BibitemShut {NoStop}%
\bibitem [{\citenamefont {Park}\ and\ \citenamefont
  {Shrout}(1997)}]{Park97:82}%
  \BibitemOpen
  \bibfield  {author} {\bibinfo {author} {\bibfnamefont {S.~E.}\ \bibnamefont
  {Park}}\ and\ \bibinfo {author} {\bibfnamefont {T.~R.}\ \bibnamefont
  {Shrout}},\ }\href@noop {} {\bibfield  {journal} {\bibinfo  {journal} {J.
  Appl. Phys.}\ }\textbf {\bibinfo {volume} {82}},\ \bibinfo {pages} {1804}
  (\bibinfo {year} {1997})}\BibitemShut {NoStop}%
\bibitem [{\citenamefont {Service}(1997)}]{Service97:275}%
  \BibitemOpen
  \bibfield  {author} {\bibinfo {author} {\bibfnamefont {R.~F.}\ \bibnamefont
  {Service}},\ }\href@noop {} {\bibfield  {journal} {\bibinfo  {journal}
  {Science}\ }\textbf {\bibinfo {volume} {275}},\ \bibinfo {pages} {1878}
  (\bibinfo {year} {1997})}\BibitemShut {NoStop}%
\bibitem [{\citenamefont {Cowley}\ \emph {et~al.}(2011)\citenamefont {Cowley},
  \citenamefont {Gvasaliya}, \citenamefont {Lushnikov}, \citenamefont
  {Roessli},\ and\ \citenamefont {Rotaru}}]{Cowley10:60}%
  \BibitemOpen
  \bibfield  {author} {\bibinfo {author} {\bibfnamefont {R.~A.}\ \bibnamefont
  {Cowley}}, \bibinfo {author} {\bibfnamefont {S.~N.}\ \bibnamefont
  {Gvasaliya}}, \bibinfo {author} {\bibfnamefont {S.~G.}\ \bibnamefont
  {Lushnikov}}, \bibinfo {author} {\bibfnamefont {B.}~\bibnamefont {Roessli}},
  \ and\ \bibinfo {author} {\bibfnamefont {G.~M.}\ \bibnamefont {Rotaru}},\
  }\href@noop {} {\bibfield  {journal} {\bibinfo  {journal} {Adv. Phys.}\
  }\textbf {\bibinfo {volume} {60}},\ \bibinfo {pages} {229} (\bibinfo {year}
  {2011})}\BibitemShut {NoStop}%
\bibitem [{\citenamefont {Burns}\ and\ \citenamefont
  {Dacol}(1983)}]{Burns09:79}%
  \BibitemOpen
  \bibfield  {author} {\bibinfo {author} {\bibfnamefont {G.}~\bibnamefont
  {Burns}}\ and\ \bibinfo {author} {\bibfnamefont {F.~H.}\ \bibnamefont
  {Dacol}},\ }\href@noop {} {\bibfield  {journal} {\bibinfo  {journal} {Solid
  State Commun.}\ }\textbf {\bibinfo {volume} {48}},\ \bibinfo {pages} {853}
  (\bibinfo {year} {1983})}\BibitemShut {NoStop}%
\bibitem [{\citenamefont {Jeong}\ \emph {et~al.}(2005)\citenamefont {Jeong},
  \citenamefont {Darling}, \citenamefont {Lee}, \citenamefont {Proffen},
  \citenamefont {Heffner}, \citenamefont {Park}, \citenamefont {Hong},
  \citenamefont {Dmowski},\ and\ \citenamefont {Egami}}]{Jeong05:94}%
  \BibitemOpen
  \bibfield  {author} {\bibinfo {author} {\bibfnamefont {I.~K.}\ \bibnamefont
  {Jeong}}, \bibinfo {author} {\bibfnamefont {T.~W.}\ \bibnamefont {Darling}},
  \bibinfo {author} {\bibfnamefont {J.~K.}\ \bibnamefont {Lee}}, \bibinfo
  {author} {\bibfnamefont {T.}~\bibnamefont {Proffen}}, \bibinfo {author}
  {\bibfnamefont {R.~H.}\ \bibnamefont {Heffner}}, \bibinfo {author}
  {\bibfnamefont {J.~S.}\ \bibnamefont {Park}}, \bibinfo {author}
  {\bibfnamefont {K.~S.}\ \bibnamefont {Hong}}, \bibinfo {author}
  {\bibfnamefont {W.}~\bibnamefont {Dmowski}}, \ and\ \bibinfo {author}
  {\bibfnamefont {T.}~\bibnamefont {Egami}},\ }\href@noop {} {\bibfield
  {journal} {\bibinfo  {journal} {Phys. Rev. Lett.}\ }\textbf {\bibinfo
  {volume} {94}},\ \bibinfo {pages} {147602} (\bibinfo {year}
  {2005})}\BibitemShut {NoStop}%
\bibitem [{\citenamefont {Pirc}\ and\ \citenamefont {Blinc}(2007)}]{Pirc07:76}%
  \BibitemOpen
  \bibfield  {author} {\bibinfo {author} {\bibfnamefont {R.}~\bibnamefont
  {Pirc}}\ and\ \bibinfo {author} {\bibfnamefont {R.}~\bibnamefont {Blinc}},\
  }\href@noop {} {\bibfield  {journal} {\bibinfo  {journal} {Phys. Rev. B}\
  }\textbf {\bibinfo {volume} {76}},\ \bibinfo {pages} {020101(R)} (\bibinfo
  {year} {2007})}\BibitemShut {NoStop}%
\bibitem [{\citenamefont {Vakhrushev}\ \emph {et~al.}(1996)\citenamefont
  {Vakhrushev}, \citenamefont {Naberezhnov}, \citenamefont {Sinha},
  \citenamefont {Feng},\ and\ \citenamefont {Egami}}]{Vak96:57}%
  \BibitemOpen
  \bibfield  {author} {\bibinfo {author} {\bibfnamefont {S.~B.}\ \bibnamefont
  {Vakhrushev}}, \bibinfo {author} {\bibfnamefont {A.}~\bibnamefont
  {Naberezhnov}}, \bibinfo {author} {\bibfnamefont {S.~K.}\ \bibnamefont
  {Sinha}}, \bibinfo {author} {\bibfnamefont {Y.~P.}\ \bibnamefont {Feng}}, \
  and\ \bibinfo {author} {\bibfnamefont {T.}~\bibnamefont {Egami}},\
  }\href@noop {} {\bibfield  {journal} {\bibinfo  {journal} {J. Phys. Chem.
  Solids}\ }\textbf {\bibinfo {volume} {57}},\ \bibinfo {pages} {1517}
  (\bibinfo {year} {1996})}\BibitemShut {NoStop}%
\bibitem [{\citenamefont {You}\ and\ \citenamefont {Zhang}(1997)}]{You97:79}%
  \BibitemOpen
  \bibfield  {author} {\bibinfo {author} {\bibfnamefont {H.}~\bibnamefont
  {You}}\ and\ \bibinfo {author} {\bibfnamefont {Q.~M.}\ \bibnamefont
  {Zhang}},\ }\href@noop {} {\bibfield  {journal} {\bibinfo  {journal} {Phys.
  Rev. Lett.}\ }\textbf {\bibinfo {volume} {79}},\ \bibinfo {pages} {3950}
  (\bibinfo {year} {1997})}\BibitemShut {NoStop}%
\bibitem [{\citenamefont {Vakhrushev}\ \emph {et~al.}(1998)\citenamefont
  {Vakhrushev}, \citenamefont {Okuneva},\ and\ \citenamefont
  {Savenko}}]{Vak98:40}%
  \BibitemOpen
  \bibfield  {author} {\bibinfo {author} {\bibfnamefont {S.~B.}\ \bibnamefont
  {Vakhrushev}}, \bibinfo {author} {\bibfnamefont {A.~N. N.~M.}\ \bibnamefont
  {Okuneva}}, \ and\ \bibinfo {author} {\bibfnamefont {B.~N.}\ \bibnamefont
  {Savenko}},\ }\href@noop {} {\bibfield  {journal} {\bibinfo  {journal} {Phys.
  Solid State}\ }\textbf {\bibinfo {volume} {40}},\ \bibinfo {pages} {1728}
  (\bibinfo {year} {1998})}\BibitemShut {NoStop}%
\bibitem [{\citenamefont {Hirota}\ \emph {et~al.}(2002)\citenamefont {Hirota},
  \citenamefont {Ye}, \citenamefont {Wakimoto}, \citenamefont {Gehring},\ and\
  \citenamefont {Shirane}}]{Hirota02:65}%
  \BibitemOpen
  \bibfield  {author} {\bibinfo {author} {\bibfnamefont {K.}~\bibnamefont
  {Hirota}}, \bibinfo {author} {\bibfnamefont {Z.~G.}\ \bibnamefont {Ye}},
  \bibinfo {author} {\bibfnamefont {S.}~\bibnamefont {Wakimoto}}, \bibinfo
  {author} {\bibfnamefont {P.~M.}\ \bibnamefont {Gehring}}, \ and\ \bibinfo
  {author} {\bibfnamefont {G.}~\bibnamefont {Shirane}},\ }\href@noop {}
  {\bibfield  {journal} {\bibinfo  {journal} {Phys. Rev. B}\ }\textbf {\bibinfo
  {volume} {65}},\ \bibinfo {pages} {104105} (\bibinfo {year}
  {2002})}\BibitemShut {NoStop}%
\bibitem [{\citenamefont {Hiraka}\ \emph {et~al.}(2004)\citenamefont {Hiraka},
  \citenamefont {Lee}, \citenamefont {Gehring}, \citenamefont {Xu},\ and\
  \citenamefont {Shirane}}]{Hiraka04:70}%
  \BibitemOpen
  \bibfield  {author} {\bibinfo {author} {\bibfnamefont {H.}~\bibnamefont
  {Hiraka}}, \bibinfo {author} {\bibfnamefont {S.~H.}\ \bibnamefont {Lee}},
  \bibinfo {author} {\bibfnamefont {P.~M.}\ \bibnamefont {Gehring}}, \bibinfo
  {author} {\bibfnamefont {G.}~\bibnamefont {Xu}}, \ and\ \bibinfo {author}
  {\bibfnamefont {G.}~\bibnamefont {Shirane}},\ }\href@noop {} {\bibfield
  {journal} {\bibinfo  {journal} {Phys. Rev. B}\ }\textbf {\bibinfo {volume}
  {70}},\ \bibinfo {pages} {184105} (\bibinfo {year} {2004})}\BibitemShut
  {NoStop}%
\bibitem [{\citenamefont {Gehring}\ \emph {et~al.}(2009)\citenamefont
  {Gehring}, \citenamefont {Hiraka}, \citenamefont {Stock}, \citenamefont
  {Lee}, \citenamefont {Chen}, \citenamefont {Ye}, \citenamefont {Vakhrushev},\
  and\ \citenamefont {Chowdhuri}}]{Gehring09:79}%
  \BibitemOpen
  \bibfield  {author} {\bibinfo {author} {\bibfnamefont {P.~M.}\ \bibnamefont
  {Gehring}}, \bibinfo {author} {\bibfnamefont {H.}~\bibnamefont {Hiraka}},
  \bibinfo {author} {\bibfnamefont {C.}~\bibnamefont {Stock}}, \bibinfo
  {author} {\bibfnamefont {S.~H.}\ \bibnamefont {Lee}}, \bibinfo {author}
  {\bibfnamefont {W.}~\bibnamefont {Chen}}, \bibinfo {author} {\bibfnamefont
  {Z.~G.}\ \bibnamefont {Ye}}, \bibinfo {author} {\bibfnamefont {S.~B.}\
  \bibnamefont {Vakhrushev}}, \ and\ \bibinfo {author} {\bibfnamefont
  {Z.}~\bibnamefont {Chowdhuri}},\ }\href@noop {} {\bibfield  {journal}
  {\bibinfo  {journal} {Phys. Rev. B}\ }\textbf {\bibinfo {volume} {79}},\
  \bibinfo {pages} {224109} (\bibinfo {year} {2009})}\BibitemShut {NoStop}%
\bibitem [{\citenamefont {Stock}\ \emph {et~al.}(2007)\citenamefont {Stock},
  \citenamefont {Xu}, \citenamefont {Gehring}, \citenamefont {Luo},
  \citenamefont {Zhao}, \citenamefont {Cao}, \citenamefont {Li}, \citenamefont
  {Viehland},\ and\ \citenamefont {Shirane}}]{Stock07:76}%
  \BibitemOpen
  \bibfield  {author} {\bibinfo {author} {\bibfnamefont {C.}~\bibnamefont
  {Stock}}, \bibinfo {author} {\bibfnamefont {G.}~\bibnamefont {Xu}}, \bibinfo
  {author} {\bibfnamefont {P.~M.}\ \bibnamefont {Gehring}}, \bibinfo {author}
  {\bibfnamefont {H.}~\bibnamefont {Luo}}, \bibinfo {author} {\bibfnamefont
  {X.}~\bibnamefont {Zhao}}, \bibinfo {author} {\bibfnamefont {H.}~\bibnamefont
  {Cao}}, \bibinfo {author} {\bibfnamefont {J.~F.}\ \bibnamefont {Li}},
  \bibinfo {author} {\bibfnamefont {D.}~\bibnamefont {Viehland}}, \ and\
  \bibinfo {author} {\bibfnamefont {G.}~\bibnamefont {Shirane}},\ }\href@noop
  {} {\bibfield  {journal} {\bibinfo  {journal} {Phys. Rev. B}\ }\textbf
  {\bibinfo {volume} {76}},\ \bibinfo {pages} {064122} (\bibinfo {year}
  {2007})}\BibitemShut {NoStop}%
\bibitem [{\citenamefont {Toulouse}(2008)}]{Toulouse08:369}%
  \BibitemOpen
  \bibfield  {author} {\bibinfo {author} {\bibfnamefont {J.}~\bibnamefont
  {Toulouse}},\ }\href@noop {} {\bibfield  {journal} {\bibinfo  {journal}
  {Ferroelectrics}\ }\textbf {\bibinfo {volume} {369}},\ \bibinfo {pages} {203}
  (\bibinfo {year} {2008})}\BibitemShut {NoStop}%
\bibitem [{\citenamefont {Fisch}(2003)}]{Fisch03:67}%
  \BibitemOpen
  \bibfield  {author} {\bibinfo {author} {\bibfnamefont {R.}~\bibnamefont
  {Fisch}},\ }\href@noop {} {\bibfield  {journal} {\bibinfo  {journal} {Phys.
  Rev. B}\ }\textbf {\bibinfo {volume} {67}},\ \bibinfo {pages} {094110}
  (\bibinfo {year} {2003})}\BibitemShut {NoStop}%
\bibitem [{\citenamefont {Pirc}\ \emph {et~al.}(2001)\citenamefont {Pirc},
  \citenamefont {Blinc},\ and\ \citenamefont {Bobnar}}]{Pirc01:63}%
  \BibitemOpen
  \bibfield  {author} {\bibinfo {author} {\bibfnamefont {R.}~\bibnamefont
  {Pirc}}, \bibinfo {author} {\bibfnamefont {R.}~\bibnamefont {Blinc}}, \ and\
  \bibinfo {author} {\bibfnamefont {V.}~\bibnamefont {Bobnar}},\ }\href@noop {}
  {\bibfield  {journal} {\bibinfo  {journal} {Phys. Rev. b}\ }\textbf {\bibinfo
  {volume} {63}},\ \bibinfo {pages} {054203} (\bibinfo {year}
  {2001})}\BibitemShut {NoStop}%
\bibitem [{\citenamefont {Westphal}\ \emph {et~al.}(1992)\citenamefont
  {Westphal}, \citenamefont {Kleemann},\ and\ \citenamefont
  {Glinchuk}}]{Westphal92:68}%
  \BibitemOpen
  \bibfield  {author} {\bibinfo {author} {\bibfnamefont {V.}~\bibnamefont
  {Westphal}}, \bibinfo {author} {\bibfnamefont {W.}~\bibnamefont {Kleemann}},
  \ and\ \bibinfo {author} {\bibfnamefont {M.~D.}\ \bibnamefont {Glinchuk}},\
  }\href@noop {} {\bibfield  {journal} {\bibinfo  {journal} {Phys. Rev. Lett.}\
  }\textbf {\bibinfo {volume} {68}},\ \bibinfo {pages} {847} (\bibinfo {year}
  {1992})}\BibitemShut {NoStop}%
\bibitem [{\citenamefont {Stock}\ \emph {et~al.}(2004)\citenamefont {Stock},
  \citenamefont {Birgeneau}, \citenamefont {Wakimoto}, \citenamefont {Gardner},
  \citenamefont {Chen}, \citenamefont {Ye},\ and\ \citenamefont
  {Shirane}}]{Stock04:69}%
  \BibitemOpen
  \bibfield  {author} {\bibinfo {author} {\bibfnamefont {C.}~\bibnamefont
  {Stock}}, \bibinfo {author} {\bibfnamefont {R.~J.}\ \bibnamefont
  {Birgeneau}}, \bibinfo {author} {\bibfnamefont {S.}~\bibnamefont {Wakimoto}},
  \bibinfo {author} {\bibfnamefont {J.~S.}\ \bibnamefont {Gardner}}, \bibinfo
  {author} {\bibfnamefont {W.}~\bibnamefont {Chen}}, \bibinfo {author}
  {\bibfnamefont {Z.~G.}\ \bibnamefont {Ye}}, \ and\ \bibinfo {author}
  {\bibfnamefont {G.}~\bibnamefont {Shirane}},\ }\href@noop {} {\bibfield
  {journal} {\bibinfo  {journal} {Phys. Rev. B}\ }\textbf {\bibinfo {volume}
  {69}},\ \bibinfo {pages} {094104} (\bibinfo {year} {2004})}\BibitemShut
  {NoStop}%
\bibitem [{\citenamefont {Swainson}\ \emph {et~al.}(2009)\citenamefont
  {Swainson}, \citenamefont {Stock}, \citenamefont {Gehring}, \citenamefont
  {Xu}, \citenamefont {Hirota}, \citenamefont {Qiu}, \citenamefont {Luo},
  \citenamefont {Zhao}, \citenamefont {Li},\ and\ \citenamefont
  {Viehland}}]{Swainson09:79}%
  \BibitemOpen
  \bibfield  {author} {\bibinfo {author} {\bibfnamefont {I.~P.}\ \bibnamefont
  {Swainson}}, \bibinfo {author} {\bibfnamefont {C.}~\bibnamefont {Stock}},
  \bibinfo {author} {\bibfnamefont {P.~M.}\ \bibnamefont {Gehring}}, \bibinfo
  {author} {\bibfnamefont {G.}~\bibnamefont {Xu}}, \bibinfo {author}
  {\bibfnamefont {K.}~\bibnamefont {Hirota}}, \bibinfo {author} {\bibfnamefont
  {Y.}~\bibnamefont {Qiu}}, \bibinfo {author} {\bibfnamefont {H.}~\bibnamefont
  {Luo}}, \bibinfo {author} {\bibfnamefont {X.}~\bibnamefont {Zhao}}, \bibinfo
  {author} {\bibfnamefont {J.~F.}\ \bibnamefont {Li}}, \ and\ \bibinfo {author}
  {\bibfnamefont {D.}~\bibnamefont {Viehland}},\ }\href@noop {} {\bibfield
  {journal} {\bibinfo  {journal} {Phys. Rev. B}\ }\textbf {\bibinfo {volume}
  {79}},\ \bibinfo {pages} {224301} (\bibinfo {year} {2009})}\BibitemShut
  {NoStop}%
\bibitem [{\citenamefont {Naberezhnov}\ \emph {et~al.}(1999)\citenamefont
  {Naberezhnov}, \citenamefont {Vakhrushev}, \citenamefont {Dorner},
  \citenamefont {Strauch},\ and\ \citenamefont {Moudden}}]{Nab99:11}%
  \BibitemOpen
  \bibfield  {author} {\bibinfo {author} {\bibfnamefont {A.}~\bibnamefont
  {Naberezhnov}}, \bibinfo {author} {\bibfnamefont {S.}~\bibnamefont
  {Vakhrushev}}, \bibinfo {author} {\bibfnamefont {B.}~\bibnamefont {Dorner}},
  \bibinfo {author} {\bibfnamefont {D.}~\bibnamefont {Strauch}}, \ and\
  \bibinfo {author} {\bibfnamefont {H.}~\bibnamefont {Moudden}},\ }\href@noop
  {} {\bibfield  {journal} {\bibinfo  {journal} {Eur. Phys. J. B}\ }\textbf
  {\bibinfo {volume} {11}},\ \bibinfo {pages} {13} (\bibinfo {year}
  {1999})}\BibitemShut {NoStop}%
\bibitem [{\citenamefont {Gehring}\ \emph {et~al.}(2001)\citenamefont
  {Gehring}, \citenamefont {Wakimoto}, \citenamefont {Ye},\ and\ \citenamefont
  {Shirane}}]{Gehring01:87}%
  \BibitemOpen
  \bibfield  {author} {\bibinfo {author} {\bibfnamefont {P.~M.}\ \bibnamefont
  {Gehring}}, \bibinfo {author} {\bibfnamefont {S.}~\bibnamefont {Wakimoto}},
  \bibinfo {author} {\bibfnamefont {Z.~G.}\ \bibnamefont {Ye}}, \ and\ \bibinfo
  {author} {\bibfnamefont {G.}~\bibnamefont {Shirane}},\ }\href@noop {}
  {\bibfield  {journal} {\bibinfo  {journal} {Phys. Rev. Lett.}\ }\textbf
  {\bibinfo {volume} {87}},\ \bibinfo {pages} {277601} (\bibinfo {year}
  {2001})}\BibitemShut {NoStop}%
\bibitem [{\citenamefont {Wakimoto}\ \emph
  {et~al.}(2002{\natexlab{a}})\citenamefont {Wakimoto}, \citenamefont {Stock},
  \citenamefont {Birgeneau}, \citenamefont {Ye}, \citenamefont {Chen},
  \citenamefont {Buyers}, \citenamefont {Gehring},\ and\ \citenamefont
  {Shirane}}]{Wakimoto02:65}%
  \BibitemOpen
  \bibfield  {author} {\bibinfo {author} {\bibfnamefont {S.}~\bibnamefont
  {Wakimoto}}, \bibinfo {author} {\bibfnamefont {C.}~\bibnamefont {Stock}},
  \bibinfo {author} {\bibfnamefont {R.~J.}\ \bibnamefont {Birgeneau}}, \bibinfo
  {author} {\bibfnamefont {Z.~G.}\ \bibnamefont {Ye}}, \bibinfo {author}
  {\bibfnamefont {W.}~\bibnamefont {Chen}}, \bibinfo {author} {\bibfnamefont
  {W.~J.~L.}\ \bibnamefont {Buyers}}, \bibinfo {author} {\bibfnamefont {P.~M.}\
  \bibnamefont {Gehring}}, \ and\ \bibinfo {author} {\bibfnamefont
  {G.}~\bibnamefont {Shirane}},\ }\href@noop {} {\bibfield  {journal} {\bibinfo
   {journal} {Phys. Rev. B}\ }\textbf {\bibinfo {volume} {65}},\ \bibinfo
  {pages} {172105} (\bibinfo {year} {2002}{\natexlab{a}})}\BibitemShut
  {NoStop}%
\bibitem [{\citenamefont {Viehland}\ \emph {et~al.}(1992)\citenamefont
  {Viehland}, \citenamefont {Jang}, \citenamefont {Cross},\ and\ \citenamefont
  {Wuttig}}]{Viehland92:46}%
  \BibitemOpen
  \bibfield  {author} {\bibinfo {author} {\bibfnamefont {D.}~\bibnamefont
  {Viehland}}, \bibinfo {author} {\bibfnamefont {S.~J.}\ \bibnamefont {Jang}},
  \bibinfo {author} {\bibfnamefont {L.~E.}\ \bibnamefont {Cross}}, \ and\
  \bibinfo {author} {\bibfnamefont {M.}~\bibnamefont {Wuttig}},\ }\href@noop {}
  {\bibfield  {journal} {\bibinfo  {journal} {Phys. Rev. B}\ }\textbf {\bibinfo
  {volume} {46}},\ \bibinfo {pages} {8003} (\bibinfo {year}
  {1992})}\BibitemShut {NoStop}%
\bibitem [{\citenamefont {Taniguchi}\ \emph {et~al.}(2011)\citenamefont
  {Taniguchi}, \citenamefont {Itoh},\ and\ \citenamefont
  {Fu}}]{Taniguchi11:42}%
  \BibitemOpen
  \bibfield  {author} {\bibinfo {author} {\bibfnamefont {H.}~\bibnamefont
  {Taniguchi}}, \bibinfo {author} {\bibfnamefont {M.}~\bibnamefont {Itoh}}, \
  and\ \bibinfo {author} {\bibfnamefont {D.}~\bibnamefont {Fu}},\ }\href@noop
  {} {\bibfield  {journal} {\bibinfo  {journal} {J. Raman Spectrosc.}\ }\textbf
  {\bibinfo {volume} {42}},\ \bibinfo {pages} {706} (\bibinfo {year}
  {2011})}\BibitemShut {NoStop}%
\bibitem [{\citenamefont {Al-Zein}\ \emph {et~al.}(2010)\citenamefont
  {Al-Zein}, \citenamefont {Hlinka}, \citenamefont {Rouquette},\ and\
  \citenamefont {Hehlen}}]{Zein10:105}%
  \BibitemOpen
  \bibfield  {author} {\bibinfo {author} {\bibfnamefont {A.}~\bibnamefont
  {Al-Zein}}, \bibinfo {author} {\bibfnamefont {J.}~\bibnamefont {Hlinka}},
  \bibinfo {author} {\bibfnamefont {J.}~\bibnamefont {Rouquette}}, \ and\
  \bibinfo {author} {\bibfnamefont {B.}~\bibnamefont {Hehlen}},\ }\href@noop {}
  {\bibfield  {journal} {\bibinfo  {journal} {Phys. Rev. Lett.}\ }\textbf
  {\bibinfo {volume} {105}},\ \bibinfo {pages} {017601} (\bibinfo {year}
  {2010})}\BibitemShut {NoStop}%
\bibitem [{\citenamefont {Hehlen}\ \emph {et~al.}(2016)\citenamefont {Hehlen},
  \citenamefont {Al-Sabbagh}, \citenamefont {Al-Zein},\ and\ \citenamefont
  {Hlinka}}]{Hehlen16:117}%
  \BibitemOpen
  \bibfield  {author} {\bibinfo {author} {\bibfnamefont {B.}~\bibnamefont
  {Hehlen}}, \bibinfo {author} {\bibfnamefont {M.}~\bibnamefont {Al-Sabbagh}},
  \bibinfo {author} {\bibfnamefont {A.}~\bibnamefont {Al-Zein}}, \ and\
  \bibinfo {author} {\bibfnamefont {J.}~\bibnamefont {Hlinka}},\ }\href@noop {}
  {\bibfield  {journal} {\bibinfo  {journal} {Phys. Rev. Lett.}\ }\textbf
  {\bibinfo {volume} {117}},\ \bibinfo {pages} {155501} (\bibinfo {year}
  {2016})}\BibitemShut {NoStop}%
\bibitem [{\citenamefont {Gehring}\ \emph {et~al.}(2000)\citenamefont
  {Gehring}, \citenamefont {Park},\ and\ \citenamefont
  {Shirane}}]{Gehring00:84}%
  \BibitemOpen
  \bibfield  {author} {\bibinfo {author} {\bibfnamefont {P.~M.}\ \bibnamefont
  {Gehring}}, \bibinfo {author} {\bibfnamefont {S.~E.}\ \bibnamefont {Park}}, \
  and\ \bibinfo {author} {\bibfnamefont {G.}~\bibnamefont {Shirane}},\
  }\href@noop {} {\bibfield  {journal} {\bibinfo  {journal} {Phys. Rev. Lett.}\
  }\textbf {\bibinfo {volume} {84}},\ \bibinfo {pages} {5216} (\bibinfo {year}
  {2000})}\BibitemShut {NoStop}%
\bibitem [{\citenamefont {Stock}\ \emph {et~al.}(2006)\citenamefont {Stock},
  \citenamefont {Ellis}, \citenamefont {Swainson}, \citenamefont {Xu},
  \citenamefont {Hiraka}, \citenamefont {Zhong}, \citenamefont {Luo},
  \citenamefont {Zhao}, \citenamefont {Viehland}, \citenamefont {Birgeneau},\
  and\ \citenamefont {Shirane}}]{Stock06:73}%
  \BibitemOpen
  \bibfield  {author} {\bibinfo {author} {\bibfnamefont {C.}~\bibnamefont
  {Stock}}, \bibinfo {author} {\bibfnamefont {D.}~\bibnamefont {Ellis}},
  \bibinfo {author} {\bibfnamefont {I.~P.}\ \bibnamefont {Swainson}}, \bibinfo
  {author} {\bibfnamefont {G.}~\bibnamefont {Xu}}, \bibinfo {author}
  {\bibfnamefont {H.}~\bibnamefont {Hiraka}}, \bibinfo {author} {\bibfnamefont
  {Z.}~\bibnamefont {Zhong}}, \bibinfo {author} {\bibfnamefont
  {H.}~\bibnamefont {Luo}}, \bibinfo {author} {\bibfnamefont {X.}~\bibnamefont
  {Zhao}}, \bibinfo {author} {\bibfnamefont {D.}~\bibnamefont {Viehland}},
  \bibinfo {author} {\bibfnamefont {R.~J.}\ \bibnamefont {Birgeneau}}, \ and\
  \bibinfo {author} {\bibfnamefont {G.}~\bibnamefont {Shirane}},\ }\href@noop
  {} {\bibfield  {journal} {\bibinfo  {journal} {Phys. Rev. B}\ }\textbf
  {\bibinfo {volume} {73}},\ \bibinfo {pages} {064107} (\bibinfo {year}
  {2006})}\BibitemShut {NoStop}%
\bibitem [{\citenamefont {Luo}\ \emph {et~al.}(200)\citenamefont {Luo},
  \citenamefont {Xu}, \citenamefont {Xu}, \citenamefont {Wang},\ and\
  \citenamefont {Yin}}]{Luo00:39}%
  \BibitemOpen
  \bibfield  {author} {\bibinfo {author} {\bibfnamefont {H.}~\bibnamefont
  {Luo}}, \bibinfo {author} {\bibfnamefont {G.}~\bibnamefont {Xu}}, \bibinfo
  {author} {\bibfnamefont {H.}~\bibnamefont {Xu}}, \bibinfo {author}
  {\bibfnamefont {P.}~\bibnamefont {Wang}}, \ and\ \bibinfo {author}
  {\bibfnamefont {Z.}~\bibnamefont {Yin}},\ }\href@noop {} {\bibfield
  {journal} {\bibinfo  {journal} {Jpn. J. Appl. Phys.}\ }\textbf {\bibinfo
  {volume} {39}},\ \bibinfo {pages} {5581} (\bibinfo {year} {200})}\BibitemShut
  {NoStop}%
\bibitem [{\citenamefont {Hlinka}\ \emph {et~al.}(2003)\citenamefont {Hlinka},
  \citenamefont {Kamba}, \citenamefont {Petzelt}, \citenamefont {Kulda},
  \citenamefont {Randall},\ and\ \citenamefont {Zhang}}]{Hlinka03:91}%
  \BibitemOpen
  \bibfield  {author} {\bibinfo {author} {\bibfnamefont {J.}~\bibnamefont
  {Hlinka}}, \bibinfo {author} {\bibfnamefont {S.}~\bibnamefont {Kamba}},
  \bibinfo {author} {\bibfnamefont {J.}~\bibnamefont {Petzelt}}, \bibinfo
  {author} {\bibfnamefont {J.}~\bibnamefont {Kulda}}, \bibinfo {author}
  {\bibfnamefont {C.~A.}\ \bibnamefont {Randall}}, \ and\ \bibinfo {author}
  {\bibfnamefont {S.~J.}\ \bibnamefont {Zhang}},\ }\href@noop {} {\bibfield
  {journal} {\bibinfo  {journal} {Phys. Rev. Lett.}\ }\textbf {\bibinfo
  {volume} {91}},\ \bibinfo {pages} {107602} (\bibinfo {year}
  {2003})}\BibitemShut {NoStop}%
\bibitem [{\citenamefont {Wakimoto}\ \emph
  {et~al.}(2002{\natexlab{b}})\citenamefont {Wakimoto}, \citenamefont {Stock},
  \citenamefont {Ye}, \citenamefont {Chen}, \citenamefont {Gehring},\ and\
  \citenamefont {Shirane}}]{Wakimoto02:66}%
  \BibitemOpen
  \bibfield  {author} {\bibinfo {author} {\bibfnamefont {S.}~\bibnamefont
  {Wakimoto}}, \bibinfo {author} {\bibfnamefont {C.}~\bibnamefont {Stock}},
  \bibinfo {author} {\bibfnamefont {Z.~G.}\ \bibnamefont {Ye}}, \bibinfo
  {author} {\bibfnamefont {W.}~\bibnamefont {Chen}}, \bibinfo {author}
  {\bibfnamefont {P.~M.}\ \bibnamefont {Gehring}}, \ and\ \bibinfo {author}
  {\bibfnamefont {G.}~\bibnamefont {Shirane}},\ }\href@noop {} {\bibfield
  {journal} {\bibinfo  {journal} {Phys. Rev. B}\ }\textbf {\bibinfo {volume}
  {66}},\ \bibinfo {pages} {224102} (\bibinfo {year}
  {2002}{\natexlab{b}})}\BibitemShut {NoStop}%
\bibitem [{\citenamefont {Stock}\ \emph {et~al.}(2005)\citenamefont {Stock},
  \citenamefont {Luo}, \citenamefont {Viehland}, \citenamefont {Li},
  \citenamefont {Swainson}, \citenamefont {Birgeneau},\ and\ \citenamefont
  {Shirane}}]{Stock05:74}%
  \BibitemOpen
  \bibfield  {author} {\bibinfo {author} {\bibfnamefont {C.}~\bibnamefont
  {Stock}}, \bibinfo {author} {\bibfnamefont {H.}~\bibnamefont {Luo}}, \bibinfo
  {author} {\bibfnamefont {D.}~\bibnamefont {Viehland}}, \bibinfo {author}
  {\bibfnamefont {J.~F.}\ \bibnamefont {Li}}, \bibinfo {author} {\bibfnamefont
  {I.~P.}\ \bibnamefont {Swainson}}, \bibinfo {author} {\bibfnamefont {R.~J.}\
  \bibnamefont {Birgeneau}}, \ and\ \bibinfo {author} {\bibfnamefont
  {G.}~\bibnamefont {Shirane}},\ }\href@noop {} {\bibfield  {journal} {\bibinfo
   {journal} {J. Phys. Soc. Jpn.}\ }\textbf {\bibinfo {volume} {74}},\ \bibinfo
  {pages} {3002} (\bibinfo {year} {2005})}\BibitemShut {NoStop}%
\bibitem [{\citenamefont {Stock}\ \emph {et~al.}(2012)\citenamefont {Stock},
  \citenamefont {Gehring}, \citenamefont {Hiraka}, \citenamefont {Swainson},
  \citenamefont {Xu}, \citenamefont {Ye}, \citenamefont {Luo}, \citenamefont
  {Li},\ and\ \citenamefont {Viehland}}]{Stock12:86}%
  \BibitemOpen
  \bibfield  {author} {\bibinfo {author} {\bibfnamefont {C.}~\bibnamefont
  {Stock}}, \bibinfo {author} {\bibfnamefont {P.~M.}\ \bibnamefont {Gehring}},
  \bibinfo {author} {\bibfnamefont {H.}~\bibnamefont {Hiraka}}, \bibinfo
  {author} {\bibfnamefont {I.}~\bibnamefont {Swainson}}, \bibinfo {author}
  {\bibfnamefont {G.}~\bibnamefont {Xu}}, \bibinfo {author} {\bibfnamefont
  {Z.~G.}\ \bibnamefont {Ye}}, \bibinfo {author} {\bibfnamefont
  {H.}~\bibnamefont {Luo}}, \bibinfo {author} {\bibfnamefont {J.~F.}\
  \bibnamefont {Li}}, \ and\ \bibinfo {author} {\bibfnamefont {D.}~\bibnamefont
  {Viehland}},\ }\href@noop {} {\bibfield  {journal} {\bibinfo  {journal}
  {Phys. Rev. B}\ }\textbf {\bibinfo {volume} {86}},\ \bibinfo {pages} {104108}
  (\bibinfo {year} {2012})}\BibitemShut {NoStop}%
\bibitem [{\citenamefont {Stock}\ \emph {et~al.}(2010)\citenamefont {Stock},
  \citenamefont {VanEijck}, \citenamefont {Fouquet}, \citenamefont {Maccarini},
  \citenamefont {Gehring}, \citenamefont {Xu}, \citenamefont {Luo},
  \citenamefont {Zhao}, \citenamefont {Li},\ and\ \citenamefont
  {Viehland}}]{Stock10:81}%
  \BibitemOpen
  \bibfield  {author} {\bibinfo {author} {\bibfnamefont {C.}~\bibnamefont
  {Stock}}, \bibinfo {author} {\bibfnamefont {L.}~\bibnamefont {VanEijck}},
  \bibinfo {author} {\bibfnamefont {P.}~\bibnamefont {Fouquet}}, \bibinfo
  {author} {\bibfnamefont {M.}~\bibnamefont {Maccarini}}, \bibinfo {author}
  {\bibfnamefont {P.~M.}\ \bibnamefont {Gehring}}, \bibinfo {author}
  {\bibfnamefont {G.}~\bibnamefont {Xu}}, \bibinfo {author} {\bibfnamefont
  {H.}~\bibnamefont {Luo}}, \bibinfo {author} {\bibfnamefont {X.}~\bibnamefont
  {Zhao}}, \bibinfo {author} {\bibfnamefont {J.~F.}\ \bibnamefont {Li}}, \ and\
  \bibinfo {author} {\bibfnamefont {D.}~\bibnamefont {Viehland}},\ }\href@noop
  {} {\bibfield  {journal} {\bibinfo  {journal} {Phys. Rev. B}\ }\textbf
  {\bibinfo {volume} {81}},\ \bibinfo {pages} {144127} (\bibinfo {year}
  {2010})}\BibitemShut {NoStop}%
\bibitem [{\citenamefont {Cowley}(1963)}]{Cowley03:68}%
  \BibitemOpen
  \bibfield  {author} {\bibinfo {author} {\bibfnamefont {R.~A.}\ \bibnamefont
  {Cowley}},\ }\href@noop {} {\bibfield  {journal} {\bibinfo  {journal} {Adv.
  Phys.}\ }\textbf {\bibinfo {volume} {12}},\ \bibinfo {pages} {421} (\bibinfo
  {year} {1963})}\BibitemShut {NoStop}%
\bibitem [{\citenamefont {Yamada}\ and\ \citenamefont
  {Shirane}(1969)}]{Yamada69:26}%
  \BibitemOpen
  \bibfield  {author} {\bibinfo {author} {\bibfnamefont {Y.}~\bibnamefont
  {Yamada}}\ and\ \bibinfo {author} {\bibfnamefont {G.}~\bibnamefont
  {Shirane}},\ }\href@noop {} {\bibfield  {journal} {\bibinfo  {journal} {J.
  Phys. Soc. Jpn.}\ }\textbf {\bibinfo {volume} {26}},\ \bibinfo {pages} {396}
  (\bibinfo {year} {1969})}\BibitemShut {NoStop}%
\bibitem [{\citenamefont {Axe}\ \emph {et~al.}(1970)\citenamefont {Axe},
  \citenamefont {Harada},\ and\ \citenamefont {Shirane}}]{Axe70:1}%
  \BibitemOpen
  \bibfield  {author} {\bibinfo {author} {\bibfnamefont {J.~D.}\ \bibnamefont
  {Axe}}, \bibinfo {author} {\bibfnamefont {J.}~\bibnamefont {Harada}}, \ and\
  \bibinfo {author} {\bibfnamefont {G.}~\bibnamefont {Shirane}},\ }\href@noop
  {} {\bibfield  {journal} {\bibinfo  {journal} {Phys. Rev. B}\ }\textbf
  {\bibinfo {volume} {1}},\ \bibinfo {pages} {1227} (\bibinfo {year}
  {1970})}\BibitemShut {NoStop}%
\bibitem [{\citenamefont {Fano}(1961)}]{Fano61:124}%
  \BibitemOpen
  \bibfield  {author} {\bibinfo {author} {\bibfnamefont {U.}~\bibnamefont
  {Fano}},\ }\href@noop {} {\bibfield  {journal} {\bibinfo  {journal} {Phys.
  Rev.}\ }\textbf {\bibinfo {volume} {124}},\ \bibinfo {pages} {1866} (\bibinfo
  {year} {1961})}\BibitemShut {NoStop}%
\bibitem [{\citenamefont {Struzhkin}\ \emph {et~al.}(1997)\citenamefont
  {Struzhkin}, \citenamefont {Goncharov}, \citenamefont {Hemley},\ and\
  \citenamefont {Mao}}]{Struzhkin97:78}%
  \BibitemOpen
  \bibfield  {author} {\bibinfo {author} {\bibfnamefont {V.~V.}\ \bibnamefont
  {Struzhkin}}, \bibinfo {author} {\bibfnamefont {A.~F.}\ \bibnamefont
  {Goncharov}}, \bibinfo {author} {\bibfnamefont {R.~J.}\ \bibnamefont
  {Hemley}}, \ and\ \bibinfo {author} {\bibfnamefont {H.~K.}\ \bibnamefont
  {Mao}},\ }\href@noop {} {\bibfield  {journal} {\bibinfo  {journal} {Phys.
  Rev. Lett.}\ }\textbf {\bibinfo {volume} {78}},\ \bibinfo {pages} {4446}
  (\bibinfo {year} {1997})}\BibitemShut {NoStop}%
\bibitem [{\citenamefont {Aoki}\ \emph {et~al.}(1996)\citenamefont {Aoki},
  \citenamefont {Yamawaki},\ and\ \citenamefont {Sakashita}}]{Aoki96:76}%
  \BibitemOpen
  \bibfield  {author} {\bibinfo {author} {\bibfnamefont {K.}~\bibnamefont
  {Aoki}}, \bibinfo {author} {\bibfnamefont {H.}~\bibnamefont {Yamawaki}}, \
  and\ \bibinfo {author} {\bibfnamefont {M.}~\bibnamefont {Sakashita}},\
  }\href@noop {} {\bibfield  {journal} {\bibinfo  {journal} {Phys. Rev. Lett.}\
  }\textbf {\bibinfo {volume} {76}},\ \bibinfo {pages} {784} (\bibinfo {year}
  {1996})}\BibitemShut {NoStop}%
\bibitem [{\citenamefont {Shirane}\ \emph {et~al.}(1970)\citenamefont
  {Shirane}, \citenamefont {Axe}, \citenamefont {Harada},\ and\ \citenamefont
  {Linz}}]{ShiraneB70:2}%
  \BibitemOpen
  \bibfield  {author} {\bibinfo {author} {\bibfnamefont {G.}~\bibnamefont
  {Shirane}}, \bibinfo {author} {\bibfnamefont {J.~D.}\ \bibnamefont {Axe}},
  \bibinfo {author} {\bibfnamefont {J.}~\bibnamefont {Harada}}, \ and\ \bibinfo
  {author} {\bibfnamefont {A.}~\bibnamefont {Linz}},\ }\href@noop {} {\bibfield
   {journal} {\bibinfo  {journal} {Phys. Rev. B}\ }\textbf {\bibinfo {volume}
  {2}},\ \bibinfo {pages} {3651} (\bibinfo {year} {1970})}\BibitemShut
  {NoStop}%
\bibitem [{\citenamefont {Ivanov}\ \emph {et~al.}(2005)\citenamefont {Ivanov},
  \citenamefont {Kozlowski}, \citenamefont {Piesiewicz}, \citenamefont
  {Stephanovich}, \citenamefont {Weron},\ and\ \citenamefont
  {Wymyslowski}}]{Ivanov05:47}%
  \BibitemOpen
  \bibfield  {author} {\bibinfo {author} {\bibfnamefont {M.~A.}\ \bibnamefont
  {Ivanov}}, \bibinfo {author} {\bibfnamefont {M.}~\bibnamefont {Kozlowski}},
  \bibinfo {author} {\bibfnamefont {T.}~\bibnamefont {Piesiewicz}}, \bibinfo
  {author} {\bibfnamefont {V.~A.}\ \bibnamefont {Stephanovich}}, \bibinfo
  {author} {\bibfnamefont {A.}~\bibnamefont {Weron}}, \ and\ \bibinfo {author}
  {\bibfnamefont {A.}~\bibnamefont {Wymyslowski}},\ }\href@noop {} {\bibfield
  {journal} {\bibinfo  {journal} {Phys. Solid State}\ }\textbf {\bibinfo
  {volume} {47}},\ \bibinfo {pages} {1928} (\bibinfo {year}
  {2005})}\BibitemShut {NoStop}%
\bibitem [{\citenamefont {Xu}\ \emph {et~al.}(2008)\citenamefont {Xu},
  \citenamefont {Wen}, \citenamefont {Stock},\ and\ \citenamefont
  {Gehring}}]{Xu08:7}%
  \BibitemOpen
  \bibfield  {author} {\bibinfo {author} {\bibfnamefont {G.}~\bibnamefont
  {Xu}}, \bibinfo {author} {\bibfnamefont {J.~S.}\ \bibnamefont {Wen}},
  \bibinfo {author} {\bibfnamefont {C.}~\bibnamefont {Stock}}, \ and\ \bibinfo
  {author} {\bibfnamefont {P.~M.}\ \bibnamefont {Gehring}},\ }\href@noop {}
  {\bibfield  {journal} {\bibinfo  {journal} {Nat. Mater.}\ }\textbf {\bibinfo
  {volume} {7}},\ \bibinfo {pages} {562} (\bibinfo {year} {2008})}\BibitemShut
  {NoStop}%
\bibitem [{\citenamefont {Schneeloch}\ \emph {et~al.}(2015)\citenamefont
  {Schneeloch}, \citenamefont {Xu}, \citenamefont {Winn}, \citenamefont
  {Stock}, \citenamefont {Gehring}, \citenamefont {Birgeneau},\ and\
  \citenamefont {Xu}}]{Schneeloch15:92}%
  \BibitemOpen
  \bibfield  {author} {\bibinfo {author} {\bibfnamefont {J.~A.}\ \bibnamefont
  {Schneeloch}}, \bibinfo {author} {\bibfnamefont {Z.~J.}\ \bibnamefont {Xu}},
  \bibinfo {author} {\bibfnamefont {B.}~\bibnamefont {Winn}}, \bibinfo {author}
  {\bibfnamefont {C.}~\bibnamefont {Stock}}, \bibinfo {author} {\bibfnamefont
  {P.~M.}\ \bibnamefont {Gehring}}, \bibinfo {author} {\bibfnamefont {R.~J.}\
  \bibnamefont {Birgeneau}}, \ and\ \bibinfo {author} {\bibfnamefont
  {G.}~\bibnamefont {Xu}},\ }\href@noop {} {\bibfield  {journal} {\bibinfo
  {journal} {Phys. Rev. B}\ }\textbf {\bibinfo {volume} {92}},\ \bibinfo
  {pages} {214302} (\bibinfo {year} {2015})}\BibitemShut {NoStop}%
\bibitem [{\citenamefont {Xu}\ \emph {et~al.}(2012)\citenamefont {Xu},
  \citenamefont {Wen}, \citenamefont {Mamontov}, \citenamefont {Stock},
  \citenamefont {Gehring},\ and\ \citenamefont {Xu}}]{Xu12:86}%
  \BibitemOpen
  \bibfield  {author} {\bibinfo {author} {\bibfnamefont {Z.}~\bibnamefont
  {Xu}}, \bibinfo {author} {\bibfnamefont {J.}~\bibnamefont {Wen}}, \bibinfo
  {author} {\bibfnamefont {E.}~\bibnamefont {Mamontov}}, \bibinfo {author}
  {\bibfnamefont {C.}~\bibnamefont {Stock}}, \bibinfo {author} {\bibfnamefont
  {P.~M.}\ \bibnamefont {Gehring}}, \ and\ \bibinfo {author} {\bibfnamefont
  {G.}~\bibnamefont {Xu}},\ }\href@noop {} {\bibfield  {journal} {\bibinfo
  {journal} {Phys. Rev. B}\ }\textbf {\bibinfo {volume} {86}},\ \bibinfo
  {pages} {144106} (\bibinfo {year} {2012})}\BibitemShut {NoStop}%
\bibitem [{\citenamefont {Blinc}\ \emph {et~al.}(2003)\citenamefont {Blinc},
  \citenamefont {Laguta},\ and\ \citenamefont {Zalar}}]{Blinc03:91}%
  \BibitemOpen
  \bibfield  {author} {\bibinfo {author} {\bibfnamefont {R.}~\bibnamefont
  {Blinc}}, \bibinfo {author} {\bibfnamefont {V.}~\bibnamefont {Laguta}}, \
  and\ \bibinfo {author} {\bibfnamefont {B.}~\bibnamefont {Zalar}},\
  }\href@noop {} {\bibfield  {journal} {\bibinfo  {journal} {Phys. Rev. Lett.}\
  }\textbf {\bibinfo {volume} {91}},\ \bibinfo {pages} {247601} (\bibinfo
  {year} {2003})}\BibitemShut {NoStop}%
\bibitem [{\citenamefont {Laguta}\ \emph {et~al.}(2003)\citenamefont {Laguta},
  \citenamefont {Glinchuk}, \citenamefont {Nokhrin}, \citenamefont {Bykov},
  \citenamefont {Blinc}, \citenamefont {Gregorovic},\ and\ \citenamefont
  {Zalar}}]{Laguta03:67}%
  \BibitemOpen
  \bibfield  {author} {\bibinfo {author} {\bibfnamefont {V.~V.}\ \bibnamefont
  {Laguta}}, \bibinfo {author} {\bibfnamefont {M.~D.}\ \bibnamefont
  {Glinchuk}}, \bibinfo {author} {\bibfnamefont {S.~N.}\ \bibnamefont
  {Nokhrin}}, \bibinfo {author} {\bibfnamefont {I.~P.}\ \bibnamefont {Bykov}},
  \bibinfo {author} {\bibfnamefont {R.}~\bibnamefont {Blinc}}, \bibinfo
  {author} {\bibfnamefont {A.}~\bibnamefont {Gregorovic}}, \ and\ \bibinfo
  {author} {\bibfnamefont {B.}~\bibnamefont {Zalar}},\ }\href@noop {}
  {\bibfield  {journal} {\bibinfo  {journal} {Phys. Rev. B}\ }\textbf {\bibinfo
  {volume} {67}},\ \bibinfo {pages} {104106} (\bibinfo {year}
  {2003})}\BibitemShut {NoStop}%
\bibitem [{\citenamefont {Bussmann-Holder}\ and\ \citenamefont
  {Bishop}(2004)}]{Buss04:16}%
  \BibitemOpen
  \bibfield  {author} {\bibinfo {author} {\bibfnamefont {A.}~\bibnamefont
  {Bussmann-Holder}}\ and\ \bibinfo {author} {\bibfnamefont {A.~R.}\
  \bibnamefont {Bishop}},\ }\href@noop {} {\bibfield  {journal} {\bibinfo
  {journal} {J. Phys. Condens. Matter}\ }\textbf {\bibinfo {volume} {16}},\
  \bibinfo {pages} {L313} (\bibinfo {year} {2004})}\BibitemShut {NoStop}%
\bibitem [{\citenamefont {Takenaka}\ \emph {et~al.}(2017)\citenamefont
  {Takenaka}, \citenamefont {Grinberg}, \citenamefont {Liu},\ and\
  \citenamefont {Rappe}}]{Takenaka17:546}%
  \BibitemOpen
  \bibfield  {author} {\bibinfo {author} {\bibfnamefont {H.}~\bibnamefont
  {Takenaka}}, \bibinfo {author} {\bibfnamefont {I.}~\bibnamefont {Grinberg}},
  \bibinfo {author} {\bibfnamefont {S.}~\bibnamefont {Liu}}, \ and\ \bibinfo
  {author} {\bibfnamefont {A.~M.}\ \bibnamefont {Rappe}},\ }\href@noop {}
  {\bibfield  {journal} {\bibinfo  {journal} {Nature}\ }\textbf {\bibinfo
  {volume} {546}},\ \bibinfo {pages} {391} (\bibinfo {year}
  {2017})}\BibitemShut {NoStop}%
\bibitem [{\citenamefont {Bosak}\ \emph {et~al.}(2012)\citenamefont {Bosak},
  \citenamefont {Chernyshov}, \citenamefont {Vakhrushev},\ and\ \citenamefont
  {Krisch}}]{Bosak12:68}%
  \BibitemOpen
  \bibfield  {author} {\bibinfo {author} {\bibfnamefont {A.}~\bibnamefont
  {Bosak}}, \bibinfo {author} {\bibfnamefont {D.}~\bibnamefont {Chernyshov}},
  \bibinfo {author} {\bibfnamefont {S.}~\bibnamefont {Vakhrushev}}, \ and\
  \bibinfo {author} {\bibfnamefont {M.}~\bibnamefont {Krisch}},\ }\href@noop {}
  {\bibfield  {journal} {\bibinfo  {journal} {Act Cryst. A}\ }\textbf {\bibinfo
  {volume} {26}},\ \bibinfo {pages} {117} (\bibinfo {year} {2012})}\BibitemShut
  {NoStop}%
\bibitem [{\citenamefont {Pitaevskii}(1959)}]{Pita59:36}%
  \BibitemOpen
  \bibfield  {author} {\bibinfo {author} {\bibfnamefont {L.~P.}\ \bibnamefont
  {Pitaevskii}},\ }\href@noop {} {\bibfield  {journal} {\bibinfo  {journal}
  {JETP}\ }\textbf {\bibinfo {volume} {9}},\ \bibinfo {pages} {830} (\bibinfo
  {year} {1959})}\BibitemShut {NoStop}%
\bibitem [{\citenamefont {Woods}\ and\ \citenamefont
  {Cowley}(1973)}]{Woods73:36}%
  \BibitemOpen
  \bibfield  {author} {\bibinfo {author} {\bibfnamefont {A.~D.~B.}\
  \bibnamefont {Woods}}\ and\ \bibinfo {author} {\bibfnamefont {R.~A.}\
  \bibnamefont {Cowley}},\ }\href@noop {} {\bibfield  {journal} {\bibinfo
  {journal} {Rep. Prog. Phys.}\ }\textbf {\bibinfo {volume} {36}},\ \bibinfo
  {pages} {1135} (\bibinfo {year} {1973})}\BibitemShut {NoStop}%
\bibitem [{\citenamefont {Suhl}(1957)}]{Shuhl57:1}%
  \BibitemOpen
  \bibfield  {author} {\bibinfo {author} {\bibfnamefont {H.}~\bibnamefont
  {Suhl}},\ }\href@noop {} {\bibfield  {journal} {\bibinfo  {journal} {J. Phys.
  Chem. Solids}\ }\textbf {\bibinfo {volume} {1}},\ \bibinfo {pages} {209}
  (\bibinfo {year} {1957})}\BibitemShut {NoStop}%
\bibitem [{\citenamefont {Orbach}(1966)}]{Orbach66:16}%
  \BibitemOpen
  \bibfield  {author} {\bibinfo {author} {\bibfnamefont {R.}~\bibnamefont
  {Orbach}},\ }\href@noop {} {\bibfield  {journal} {\bibinfo  {journal} {Phys.
  Rev. Lett.}\ }\textbf {\bibinfo {volume} {16}},\ \bibinfo {pages} {15}
  (\bibinfo {year} {1966})}\BibitemShut {NoStop}%
\bibitem [{\citenamefont {Klemens}(1967)}]{Klemens67:28}%
  \BibitemOpen
  \bibfield  {author} {\bibinfo {author} {\bibfnamefont {P.~G.}\ \bibnamefont
  {Klemens}},\ }\href@noop {} {\bibfield  {journal} {\bibinfo  {journal} {J.
  Appl. Phys.}\ }\textbf {\bibinfo {volume} {38}},\ \bibinfo {pages} {4573}
  (\bibinfo {year} {1967})}\BibitemShut {NoStop}%
\bibitem [{\citenamefont {Klemens}(1966)}]{Klemens66:148}%
  \BibitemOpen
  \bibfield  {author} {\bibinfo {author} {\bibfnamefont {P.~G.}\ \bibnamefont
  {Klemens}},\ }\href@noop {} {\bibfield  {journal} {\bibinfo  {journal} {Phys.
  Rev.}\ }\textbf {\bibinfo {volume} {148}},\ \bibinfo {pages} {845} (\bibinfo
  {year} {1966})}\BibitemShut {NoStop}%
\bibitem [{\citenamefont {Lax}\ \emph {et~al.}(1981)\citenamefont {Lax},
  \citenamefont {Hu},\ and\ \citenamefont {Narayanamurti}}]{Lax81:23}%
  \BibitemOpen
  \bibfield  {author} {\bibinfo {author} {\bibfnamefont {M.}~\bibnamefont
  {Lax}}, \bibinfo {author} {\bibfnamefont {P.}~\bibnamefont {Hu}}, \ and\
  \bibinfo {author} {\bibfnamefont {V.}~\bibnamefont {Narayanamurti}},\
  }\href@noop {} {\bibfield  {journal} {\bibinfo  {journal} {Phys. Rev. B}\
  }\textbf {\bibinfo {volume} {23}},\ \bibinfo {pages} {3095} (\bibinfo {year}
  {1981})}\BibitemShut {NoStop}%
\bibitem [{\citenamefont {Bose}\ \emph {et~al.}(1999)\citenamefont {Bose},
  \citenamefont {Kirkpatrick},\ and\ \citenamefont {Dennis}}]{Bose99:271}%
  \BibitemOpen
  \bibfield  {author} {\bibinfo {author} {\bibfnamefont {S.~K.}\ \bibnamefont
  {Bose}}, \bibinfo {author} {\bibfnamefont {S.~M.}\ \bibnamefont
  {Kirkpatrick}}, \ and\ \bibinfo {author} {\bibfnamefont {W.~M.}\ \bibnamefont
  {Dennis}},\ }\href@noop {} {\bibfield  {journal} {\bibinfo  {journal}
  {Physica B}\ }\textbf {\bibinfo {volume} {271}},\ \bibinfo {pages} {198}
  (\bibinfo {year} {1999})}\BibitemShut {NoStop}%
\bibitem [{\citenamefont {Cao}\ \emph {et~al.}(2008)\citenamefont {Cao},
  \citenamefont {Stock}, \citenamefont {Xu}, \citenamefont {Gehring},
  \citenamefont {Li},\ and\ \citenamefont {Viehland}}]{Cao08:78}%
  \BibitemOpen
  \bibfield  {author} {\bibinfo {author} {\bibfnamefont {H.}~\bibnamefont
  {Cao}}, \bibinfo {author} {\bibfnamefont {C.}~\bibnamefont {Stock}}, \bibinfo
  {author} {\bibfnamefont {G.}~\bibnamefont {Xu}}, \bibinfo {author}
  {\bibfnamefont {P.~M.}\ \bibnamefont {Gehring}}, \bibinfo {author}
  {\bibfnamefont {J.}~\bibnamefont {Li}}, \ and\ \bibinfo {author}
  {\bibfnamefont {D.}~\bibnamefont {Viehland}},\ }\href@noop {} {\bibfield
  {journal} {\bibinfo  {journal} {Phys. Rev. B}\ }\textbf {\bibinfo {volume}
  {78}},\ \bibinfo {pages} {104103} (\bibinfo {year} {2008})}\BibitemShut
  {NoStop}%
\bibitem [{\citenamefont {La-Orauttapong}\ \emph {et~al.}(2002)\citenamefont
  {La-Orauttapong}, \citenamefont {Noheda}, \citenamefont {Ye}, \citenamefont
  {Gehring}, \citenamefont {Toulouse}, \citenamefont {Cox},\ and\ \citenamefont
  {Shirane}}]{Gop02:65}%
  \BibitemOpen
  \bibfield  {author} {\bibinfo {author} {\bibfnamefont {D.}~\bibnamefont
  {La-Orauttapong}}, \bibinfo {author} {\bibfnamefont {B.}~\bibnamefont
  {Noheda}}, \bibinfo {author} {\bibfnamefont {Z.~G.}\ \bibnamefont {Ye}},
  \bibinfo {author} {\bibfnamefont {P.~M.}\ \bibnamefont {Gehring}}, \bibinfo
  {author} {\bibfnamefont {J.}~\bibnamefont {Toulouse}}, \bibinfo {author}
  {\bibfnamefont {D.~E.}\ \bibnamefont {Cox}}, \ and\ \bibinfo {author}
  {\bibfnamefont {G.}~\bibnamefont {Shirane}},\ }\href@noop {} {\bibfield
  {journal} {\bibinfo  {journal} {Phys. Rev. B}\ }\textbf {\bibinfo {volume}
  {65}},\ \bibinfo {pages} {144101} (\bibinfo {year} {2002})}\BibitemShut
  {NoStop}%
\end{thebibliography}

%

\end{document}